\documentclass[british]{article}
\usepackage[left=1in,top=1in,right=1in,bottom=1in,nohead,paperwidth=8.5in, paperheight=11in]{geometry} %1 inch margins, 8 1/2 x 11-inch pages: standard choices for most journals\
\usepackage{amsmath}
\usepackage{stmaryrd}
\usepackage{natbib} %standard package for references
\usepackage{float}
\usepackage{rotating}
\usepackage{multirow}
\usepackage{algorithm}
\usepackage[titletoc, title]{appendix}
\usepackage{capt-of}
\usepackage{adjustbox}
\usepackage{graphicx}
\usepackage{lscape}
\usepackage{caption}
\usepackage{subcaption}
\usepackage{enumerate}
\usepackage{color}
\usepackage{lineno}
\usepackage{threeparttable}

\graphicspath{{./Figures/}}

\title{\bf Commodity Dynamics: A Sparse Multi-class Approach}
\author{Luca Barbaglia\footnote{Corresponding author: KU Leuven, Faculty of Economics and Business, Naamsestraat 69, B-3000
		Leuven, Belgium. \hspace{2cm}
		\textit{Email address}: luca.barbaglia@kuleuven.be, \textit{Phone}: +32 16 37 37 84.}, Ines Wilms, and Christophe Croux \\ \textit{\small KU Leuven, Faculty of Economics and Business}}
\date{}

\begin{document}
	\maketitle
	
	{\bf Abstract.}
		The correct understanding of commodity price dynamics can bring relevant improvements in terms of policy formulation both for developing and developed countries.  
		% Sparsity 
		Agricultural, metal and energy commodity prices might depend on each other: although we expect few important effects among the total number of possible ones, some price effects among different commodities might still be substantial.
		% Multi-class 
		Moreover, the increasing integration of the world economy suggests that these effects should be comparable for different markets. 
		% Our method 
		This paper introduces a sparse estimator of the Multi-class Vector AutoRegressive model to detect common price effects between a large number of commodities, for different markets or investment portfolios. 
		% Results 
		In a first application, we consider agricultural, metal and energy commodities for three different markets. We show a large prevalence of effects involving metal commodities in the Chinese and Indian markets, and the existence of asymmetric price effects. In a second application, we analyze commodity prices for five different investment portfolios, and highlight the existence of important effects from energy to agricultural commodities. The relevance of biofuels is hereby confirmed. Overall, we find stronger similarities in commodity price effects among portfolios than among markets. \\
	
	\bigskip
	
	%{\bf Keywords:} Commodity prices, Fused Lasso, Multi-class estimation, Vector AutoRegressive Model
	

\newpage
%\linenumbers

\section{Introduction \label{intro}}
Commodity price dynamics are crucial for the worldwide economic activity \citep{Labys06}. Up to 20\% of the world merchandise trade involves commodities \citep{Nazlioglu13}.  Commodities are central for both producing and consuming countries. 
For producing countries,  export earnings from commodities are often the main source of revenues. Therefore, commodity price dynamics have an important impact on the macro-economic performance and living standards of these countries, often developing countries \citep{Cashin02, Deaton99, Rossen15}. 
For consuming countries,  commodities are important inputs for many industries. 
As such, understanding commodity price dynamics is essential for economic planning and forecasting \citep{Arezki14, Rossen15}.
\textcolor{black}{Hence, studying commodity price dynamics has social and political relevance.}

% Use of VAR and overparametrization
 We study the effects between a large number of commodity prices. The Vector AutoRegressive (VAR) model is a standard tool to model these commodity price dynamics \citep{Akram09, Rezitis15, Smiech15}.  Most studies (e.g. \citealp{Akram09,Rezitis15,Smiech15}) focus on one type of commodities, such as agriculture, energy, or metal, and model the price effects between a relatively limited number of commodities. They do this to tackle the over-parametrization problem of the VAR model since the number of parameters that need to be estimated increases quadratically with the number of included time series. An exception is \cite{Rossen15} who models up to 20 commodities, but requires more than 100 years of monthly data to estimate the model using the standard least-squares estimator.

% Contribution 1: commodity types --> sparsity
We contribute to the extant literature on commodities by modeling a large set of prices belonging to different commodity types - namely \textit{agriculture}, \textit{energy} and \textit{metal} - in a VAR framework.  From an economic point of view, commodity prices can be interlinked. Commodities are exposed to spillover effects, that is price effects from one commodity type (e.g. energy) to another (e.g. agriculture), as their production and consumption might be dependent on each other \citep{Akram09,Nazlioglu13, Smiech15}. Still, most studies consider only the effects between either energy and/or agricultural goods \citep{Balcombe08, Chen10, Hassouneh12, Nazlioglu12, Serra11mean}. On the contrary, only few studies, such as \cite{Chen15}, jointly study the effects between energy, metal and agricultural commodities. We extend this branch of the literature and jointly model the price effects between agriculture, energy and metal commodities. Although price effects among these commodities might be substantial, we do not expect that each commodity influences each and every other commodity. To detect the most important commodity price effects, we use \textit{sparse} estimation. The estimation is sparse in the sense that some of the commodity price effects are estimated as exactly zero. As such, only a small number of effects are estimated as non-zero, which eases the interpretation.

% Contribution 2: markets --> multi-class
Another relevant contribution to the literature on commodities is that we jointly model commodity price effects for different \textit{markets}. We would expect commodity price effects to be similar for the different markets since commodity prices are assumed to follow global macro-economic trends (\citealp{Rossen15, Smiech15} and references therein). In the economic literature, the standard methodology requires to first study each market independently, and only in a second step to verify the existence of comparable price effects for the different markets. For instance, \cite{Rapsomanikis11} analyzes the price transmission structure between a set of six food commodity markets in developing countries and the world index: he models each market separately and then compares the results in order to identify similar patterns. We extend this approach and propose, instead of a standard VAR model, a \textit{Multi-class VAR}, where the different classes are the different markets. By jointly studying several markets in one large model we expect to obtain a more accurate estimation of the commodity effects and a more reliable comparison of these effects for different markets.

% Network analysis]
We also propose a network analysis tool for the interpretation of the estimated effects that yields insightful results for commodity analysts. \cite{Yang00} were the first to provide a network representation of commodity effects, but no further attempts followed. Nevertheless, network analysis has experienced great popularity in recent years in the field of financial econometrics \citep{Diebold15} and big-data analysis \citep{Kolaczyk09}. Our approach consists in drawing a network, one for each market, where only the most important commodity price effects are visualized. Similarities and/or differences in price effects can be immediately detected by comparing the different market networks.

% Results
We employ the sparse estimator of the Multi-class VAR to verify the interactions in two data sets of commodity spot prices: a \textit{market} and a \textit{portfolio} data set.
The first data set considers prices of $J=14$ energy, metal and agricultural commodities for each of the $K=3$ different markets, namely World, India and China.
Our results highlight the existence of effects from energy towards agriculture in all markets, and from energy and agriculture towards metal in the Indian and Chinese market. 
The second data set considers prices of $J=17$ global, energy, metal and agricultural commodities for each of the $K=5$ investment portfolios. 
As for the markets, we expect to find comparable price effects for the portfolios \citep{Anson06}.
We detect important effects from energy towards agricultural commodities, in line with \cite{Chen10, Nazlioglu12, Rezitis15}, and towards precious metals, in line with \cite{Sari10}.

% Structure
The remainder of this article is organized as follows. The next section reviews the recent literature on commodity price dynamics. Section \ref{method} introduces the Multi-class VAR model, the corresponding sparse estimator and the network tools used to interpret the results. Section \ref{Application} considers the two data sets and verifies the effects among commodity prices using network analysis. Conclusions and directions for future research are given in Section \ref{Discussion}. 

\section{Background on Commodity Price Dynamics \label{literature}}
% Co-movements: macro-economic trends
Over the last decades commodity prices have tended to move together. A common explanation of this tendency is that commodity prices jointly respond to macro-economic trends \citep{Pindyck90}. 
Subsequent studies confirm these findings and suggest some possible economic explanations. \cite{Franckel10}, for instance, underline that rapid global growth, permissive monetary policy resulting in low interest rates, and increasing financial speculation might explain why commodity prices move closely together. Nevertheless, commodity dynamics cannot only be explained by macro-economic factors. Additional complexities might be encountered when considering, for instance, (i) different price types (e.g. spot prices or derivatives), (ii) different commodity types (e.g. energy goods or precious metals) or (iii) inventory data. Furthermore, the recent ``financialization" of commodity markets, i.e. commodities treated as a distinct asset class by financial agents,  has introduced a further level of complexity \citep{Arezki14, Belke13}. In recent years, a large flow of investment has been directed towards commodities, which are considered as mere alternative investment assets by financial agents. 

% Market integration
An important branch of the commodity literature looks at price convergence between national and international markets \citep{Bukenya05}. Two theoretical frameworks emerge from the literature: the ``law of one price" and the concept of ``market integration". On the one hand, the ``law of one price" states that for a given good, should all prices be expressed in the same currency, then a single price would be present throughout the world \citep{Isard77}. On the other hand, ``market integration" is observed when related goods follow comparable patterns over markets that are differently located \citep{Ravallion86}. Both theories have been tested empirically and contradictory results have been found for commodity markets. For instance, \cite{Yang00} finds evidence of price convergence in the markets of soybean meal, whereas \cite{Bukenya05} do not verify price convergence when considering a lager set of metal and agricultural goods. 

% Developing countries
The rising importance of numerous developing countries requires special attention when studying price convergence. Indeed, the rapid growth of the BRIC has played a central role in reshaping commodity dynamics \citep{Franckel10}. The demand of energy and raw material is rising steadily driven by the growing consumption in developing countries \citep{Arezki14}. At the same time, the offer side is characterized by continuous changes and countries like India and China are now key players in the commodity market. Consider, for instance, metal commodities: China not only represents the world's largest consumer of minerals and metals, but also the world's largest producer \citep{Klotz14}.
These overall changes in demand and offer sides should increase the interconnectedness among markets and might result in stronger similarities among commodity markets \citep{Chen15}. 
To address the topic of market integration in developing countries, we compare commodity price effects for the Chinese, Indian and World markets.

% Technological changes and biofeuls 
Another critical factor that has to be taken into account when studying commodity dynamics is technological change. A technological breakthrough might drastically change commodity price dynamics. For example, the ``shale gas revolution" \citep{Arezki14}, with the development of two key drilling technologies (i.e. horizontal drilling and hydraulic fracturing), has made the extraction of unconventional gas possible. As a result, the availability of natural gas steadily expanded and the overall energy price dynamics adapted to the change. Biofuels (like ethanol or biodiesel) might be analyzed in a similar perspective since their rapid development has drastically changed the dynamics between energy and agricultural commodities. The crops employed in biofuel distillation were no longer harvested only for food or feedstock purposes, but also for energy production. While prior to 2005 there were few effects between energy and agricultural commodities, \cite{Tyner10} shows that since 2006 stronger effects have been established among energy and agricultural goods.  

% Spillover energy-agriculture
Numerous recent works have investigated the impact of biofuels on commodity dynamics, providing evidence of important effects from energy towards agricultural prices \citep{Balcombe08, Chen10, Hassouneh12, Nazlioglu12, Serra11mean}. Among them, \cite{Serra11mean}  prove the existence of strong effects between energy prices and corn in the US. Moreover, their findings confirm that the effects from energy towards corn occur through ethanol. \cite{Nazlioglu12} are the first ones to study a large set of commodities at world level. They model up to twenty agricultural prices and verify the existence of strong effects from the world oil prices to agricultural goods. 
This paper is in line with the analysis of \cite{Nazlioglu12} and investigates the effects from energy towards agricultural commodities and vice versa, for different markets and investment portfolios.

\section{Sparse Estimation of Multi-class VAR Models \label{method}}

\subsection{Model specification}
To study the commodity dynamics between $J$ commodities for $K$ markets (or $K$ portfolios), we use the Multi-class VAR model of order $P$ with $K$ classes and  $J$ time series  given by

\begin{equation}
\textbf{y}_{t}^{(k)}=\textbf{B}_{1}^{(k)}\textbf{y}^{(k)}_{t-1}+\ldots+\textbf{B}^{(k)}_{P}\textbf{y}^{(k)}_{t-P}+\textbf{e}^{(k)}_{t}\label{eq:VAR(P)}.
\end{equation}
Here, $\textbf{y}_{t}^{(k)}=[y_{t,1}^{(k)},\ldots,y_{t,J}^{(k)}]'$ 
contains the values of the $J$ commodity prices for each class $k\in\{1,\ldots,K\}$ at a given point in time $t\in\{1,\ldots,T\}$, where $T$ is the length of the time series. 
The parameters $\textbf{B}^{(k)}_{p}$, for $p\in\{1,\ldots,P\}$ and $k\in\{1,\ldots,K\}$,  are $J\times J$ matrices
capturing commodity price effects at lag $p$ for class $k$. 
We denote each $ij^{th}$ entry of $\textbf{B}^{(k)}_{p}$ as $B^{(k)}_{p,ij}$ with $i,j \in \{1, \ldots, J\}$.
These elements measure the direct, lagged effect of one commodity on another commodity (including itself).
The error term $\textbf{e}_{t}^{(k)}$ is assumed to follow
a multivariate normal distribution $N({\bf 0}, \boldsymbol\Sigma^{(k)})$, with the
inverse error covariance matrix of each class $k$ denoted as $\boldsymbol \Omega^{(k)}=(\boldsymbol{\Sigma}^{(k)})^{-1}$. 
Without loss of generality, we assume that all series are mean centered such that no intercept is included.

In our \textit{market} application with $K=3$ markets and $J=14$ commodities, $3\times14 \times 14=588$ autoregressive parameters need to be estimated in the VAR of order $P=1$. To make estimation possible and to ease interpretation, we use sparse estimation such that many elements of  the matrices $\textbf{B}^{(k)}_{p}$, for $p\in\{1,\ldots,P\}$ and $k\in\{1,\ldots,K\}$, are estimated as zero. 
Moreover, we expect the parameter vector to be sparse in a structured manner, not in an irregular way \citep{Wainwright14}. 
In particular, we expect the effect of commodity A on commodity B to be similar for the different markets (or portfolios). Structured sparsity methods take this into account in the estimation procedure (e.g. \citealp{Jenatton11}). We use the fused lasso \citep{Tibshirani05} to encourage such shared patterns of sparsity among classes (cfr. Subsection \ref{Sparse_VAR_subsection}).

Furthermore, $3\times(14\times15)/2=315$ unique elements in the inverse error covariance matrices $\boldsymbol \Omega^{(k)}$, for $k\in\{1,\ldots,3\}$ of our \textit{market} application  need to be estimated. The elements of $\boldsymbol \Omega^{(k)}$ are the partial correlations between the error terms  of the $J$ equations of class $k$. Should the $ij^{th}$ element of $\boldsymbol \Omega^{(k)}$ be zero, then there would be no partial correlation between the error terms of equation $i$ and $j$ of class $k$. We encourage a similar sparsity pattern of the partial error correlations among classes by using the fused lasso.

\subsection{Sparse estimation of the Multi-class VAR model\label{Sparse_VAR_subsection}}
We define the penalized least-squares estimator for the Multi-class VAR. 
Let $N=T-P$ be the number of time observations actually available given the VAR of order $P$.  
We seek $\widehat{\textbf{B}}=[\widehat{\textbf{B}}^{(1)\prime}, \ldots, \widehat{\textbf{B}}^{(K)\prime}]'$ such that
\begin{multline}
\mathrm{\widehat{\boldsymbol{B}}=}\underset{B}{\operatorname{argmin}}
\sum_{k=1}^{K}\sum_{t=1}^{N}(\mathbf{y}^{(k)}_{t}-\boldsymbol{B}^{(k)}\mathbf{Y}^{(k)})'(\mathbf{y}^{(k)}_{t}-\boldsymbol{B}^{(k)}\mathbf{Y}^{(k)})\\
+\lambda_{1}\sum_{k=1}^{K}\sum_{i,j=1}^{J}\sum_{p=1}^{P}|B_{p,ij}^{(k)}|+\lambda_{2}\sum_{k\neq k'}^{K}\sum_{i,j=1}^{J}\sum_{p=1}^{P}|B_{p,ij}^{(k)}-B_{p,ij}^{(k')}|\label{eq:Intermediate_GFLasso},
\end{multline}
where $\textbf{B}^{(k)}=[\textbf{B}^{(k)\prime}_{1}, \ldots, \textbf{B}^{(k)\prime}_{P}]$ is the $J \times JP$ matrix of autoregressive coefficients, and
$\boldsymbol{Y}^{(k)}=[\textbf{y}^{(k)\prime}_{t-1}, \ldots, \textbf{y}^{(k)\prime}_{t-P}]'$ is the $JP \times 1$ vector containing the lagged time series.
Furthermore, $\lambda_{1}, \ \lambda_{2}>0$ respectively are the regularization parameters for the \textit{lasso} and \textit{fusion} penalties of the autoregressive coefficients.
In the context of univariate regression a similar estimator was proposed by \cite{Kim09}.

The first penalty corresponds to the lasso penalty \citep{Tibshirani96}: the $L_{1}$-norm of the autoregressive coefficients shrinks the elements of $\boldsymbol{B}$ and sets some of them  exactly to zero. The larger the parameter $\lambda_{1}$, the sparser the estimate of $\boldsymbol{B}$. 
The second penalty term corresponds to the fusion penalty \citep{Tibshirani05}: the $L_{1}$-norm of the difference between $B_{p,ij}^{(k)}$ and $B_{p,ij}^{(k')}$ encourages corresponding autoregressive coefficients in class $k$ and $k'$ to take the same value by shrinking their difference towards zero. The larger the value of $\lambda_2$, the more elements of $\boldsymbol{B}$ will be identical for the different classes. 

The error terms of class $k$ are not likely to be independently drawn and gains in estimation accuracy might be achieved by accounting for their correlation structure (e.g. \citealp{Rothman10}). Therefore, we use a \textit{generalized} version of the penalized least-squares criterion in \eqref{eq:Intermediate_GFLasso}. We replace the first term in formula \eqref{eq:Intermediate_GFLasso} by 
\begin{equation}
\small \sum_{k=1}^{K}\sum_{t=1}^{N}((\mathbf{y}^{(k)}_{t}-\boldsymbol{B}^{(k)}\mathbf{Y}^{(k)})'\mathbf{{\Omega}}^{(k)}(\mathbf{y}^{(k)}_{t}-\boldsymbol{B}^{(k)}\mathbf{Y}^{(k)})-\log |\mathbf{\Omega}^{(k)}|), \nonumber
\end{equation}
and also estimate the inverse error covariance matrix  $\boldsymbol{\Omega}^{(k)}$ for each class $k$. 
Moreover, we include two regularization parameters for the sparse multi-class estimation of the elements of the inverse error covariance matrix as in \cite{Danaher14}. As done for the autoregressive coefficients, we guarantee sparsity of the inverse error covariance matrix with a lasso penalty on the elements of $\boldsymbol{\Omega}^{(k)}$. 
Furthermore, we encourage similar sparsity patterns among classes with a fusion penalty, which considers the difference between corresponding elements of the inverse error covariance matrix among classes. 
Throughout the remainder of the paper we use the generalized version of the penalized least-squares criterion in \eqref{eq:Intermediate_GFLasso}. 

We use an iterative algorithm to obtain the estimates of the autoregressive parameters $\widehat{\textbf{B}}^{(k)}$ and the estimates of the inverse error covariance matrices $\widehat{\boldsymbol{\Omega}}^{(k)}$ of each class $k\in\{1,\ldots,K\}$.  
First, we solve for the autoregressive parameters conditional on the inverse error covariance matrices
using the Smoothing Proximal Gradient (SPG) algorithm (see \citealp{Chen12}). The SPG algorithm provides a flexible way to approximate the minimizer of an objective function that incorporates both fusion and lasso penalties.  
Second, we solve for the inverse error covariance matrices conditional on the autoregressive parameters, which corresponds to the Joint Graphical Lasso  on the residuals of the Multi-class VAR model. This Joint Graphical Lasso is computed using the Alternating Directions Method of Multipliers algorithm \citep{Danaher14}. The code of the algorithm is available from the authors upon request.

\subsection{Network Interpretation \label{Network_interpretation}}
The interpretation of the estimated coefficients $\widehat{{\bf B}}^{(1)},\ldots,\widehat{{\bf B}}^{(K)}$ is done via network analysis.
We focus on the cross-commodity effects\footnote{We refer to them as commodity price effects or, more simply, just effects in the remainder of the paper.}, i.e. the off-diagonal elements of $\widehat{{\bf B}}^{(1)},\ldots,\widehat{{\bf B}}^{(K)}$. We build a \textit{directed} network of commodities for each class, where the nodes are the different commodities and an edge from commodity $i$ to commodity $j$ is drawn when the corresponding price effect is estimated non-zero by the proposed estimator. 

This network analysis helps commodity analysts to interpret the results.
The effects are visualized for each class, thereby immediately highlighting the differences between classes (i.e. markets or portfolios). Only the important  effects (i.e. estimated non-zero effects) are drawn such that they are immediately distinguished from the unimportant effects (i.e. estimated zero effects). 
Measures of connectedness among commodities can be calculated from the networks. Define the following indicator of connectedness from commodity $i$ to $j$, which takes value one if the sum of the associated autoregressive coefficients is non-zero:

\[
(i\rightarrow j)=\begin{cases}
1 & if\ \sum_{p=1}^{P}\widehat{B}^{(k)}_{p,ji} \neq 0\\
0 & otherwise
\end{cases} \ \ \ \  \text{with}\ k=1, \ldots, K.
\]

Following \cite{Billio12}, we assess the connectedness of commodity $i$ by accounting for the proportion of edges (i) directed towards it (\textit{in-going} connectedness), (ii) originating from it (\textit{out-going} connectedness) and (iii) in both directions (\textit{total} connectedness). 
Each measure is standardized by the maximum value registered for class $k$ to ease the comparison.
Let $S^{(k)}$ be the set of all commodities in class $k$, we propose three measures of connectedness for commodity $i$:
\begin{enumerate}[(i)]
                \item \textit{in-going} connectedness: 
                \[
                (S^{(k)} \rightarrow i) = \frac{\nu^{(k)}_{i}}{\nu^{(k)}_{\text{max}}},
                \]
                where $\nu^{(k)}_{i}=\sum_{j \neq i}^{J} (j \rightarrow i)$ and $\nu^{(k)}_{\text{max}}=\max_{i}[\nu^{(k)}_{i}]$;

                \item \textit{out-going} connectedness: 
                \[
                (i \rightarrow S^{(k)}) = \frac{\eta^{(k)}_{i}}{\eta^{(k)}_{\text{max}}},
                \]
                where $\eta^{(k)}_{i}=\sum_{j \neq i}^{J} (i \rightarrow j)$ and $\eta^{(k)}_{\text{max}}=\max_{i}[\eta^{(k)}_{i}]$;

                \item \textit{total} connectedness: 
                               \[
                               (i \leftrightarrow S^{(k)}) = \frac{\mu^{(k)}_{i}}{\mu^{(k)}_{\text{max}}},
                               \]
                               where $\mu^{(k)}_{i}=\sum_{j \neq i}^{J} ((i \rightarrow j)+(j \rightarrow i))$ and $\mu^{(k)}_{\text{max}}=\max_{i}[\mu^{(k)}_{i}]$.

\end{enumerate}

In Section \ref{Application} we present these networks for the market and portfolio data sets using a Multi-class VAR(1) model. Then we can consider a \textit{weighted} network where the edge width represents the magnitude of the effect. Moreover, positive effects (in blue) are distinguished from negative ones (in red)\footnote{On a gray scale: positive effects are shown in dark gray and negative ones in light gray.}.
We complete the network analysis by verifying the prevalence of within and spillover effects among different commodity types. Within effects are effects among commodities of the same type, for instance agriculture. Spillover effects are effects among commodities of different types, for instance energy and agriculture.

\section{Data and Results \label{Application}}
We use the sparse estimator of the Multi-class VAR to study commodity price effects. We consider two data sets: a \textit{market} and a \textit{portfolio} data set. 
In the first one, we study spot prices of $J=14$  energy, metal and agricultural commodities for each of the $K=3$ markets.
The second data set considers indexes\footnote{The corresponding indexes from the different portfolios are all based on the same commodity spot price, but they differ from one portfolio to another since each portfolio uses a different weighting scheme. An introduction to commodity indexes is provided in \cite{Anson06}.} of spot prices of $J=17$ global, energy, metal and agricultural commodities for each of the $K=5$ portfolios. 
The choice of considering spot prices is coherent with the analysis of \cite{Pindyck01}, who underlines the existence of two parallel trade levels: a cash trading for commodity spot purchases and sales, and a storage trading for inventories held by traders and producers. These two trade levels are highly connected since a spot purchase of a commodity involves the physical delivery of the good and therefore its stocking as inventory\footnote{Electricity represents an exception since it can be hardly stocked.  For this reason we do not include electricity.}. Spot price dynamics depend on the storage levels, consequently they are less volatile than the associated derivative products, as futures or swaps \citep{Pindyck01}: this justifies our choice of using spot prices.

The details of the two data sets are respectively reported in Table \ref{Data_Set_Countries} and \ref{Data_Set_Indices}. Data are obtained from Datastream. 
The commodity spot price time series are non-stationary, we therefore follow standard practice in taking each time series in log-differences, which corresponds to the spot daily return.
Next, we standardize these spot daily returns by subtracting the mean and dividing by the standard deviation\footnote{Standardization guarantees that all time series are on the same scale, such that the penalty terms have the same impact on each time series independently of its scale \citep{Tibshirani12}.}. We check for stationarity using the Augmented Dickey-Fuller test and the pooled unit root test of \cite{Levin02}: in both cases we find strong evidence in favor of stationarity ($p$-values$<0.01$). The order $P$ of the Multi-class VAR is selected using the Bayesian Information Criterion (BIC) and equals one for both data sets. 

%\begin{savenotes}
\begin{table}
\resizebox{\textwidth}{!}{%  
\begin{threeparttable}
	\caption{\normalsize Market data set. Spot prices of $J=14$ commodities are collected for each of the $K=3$ markets. 
		 \label{Data_Set_Countries} \vspace{-0.3cm}}
	\centering

		\setlength{\tabcolsep}{0.3cm}	
		\begin{tabular}{lllll}
			\hline 
			\textbf{Commodity (type)} & \textbf{Label} & \multicolumn{3}{c}{\textbf{Source in Datastream}}\tabularnewline
						\hline
						\hline 
						\multicolumn{2}{c}{}  & \textit{World} & \textit{India} & \textit{China}\tabularnewline
						\hline 
				Crude Oil\tnote{a} \  (Energy)	&  CRUO & Crude Oil Dated FOB & Crude Oil Dated FOB & Crude Oil Dated FOB \\ 	
	 			Natural Gas\tnote{a} \  (Energy) & NATG & Henry Hub & Henry Hub &  Henry Hub
				\\
				Aluminum (Metal) & ALLU & LME 99.7\% Cash & MCI Mumbai & Bonded Whse Premium Spt\tabularnewline
				Copper (Metal) & {COPP} & {LME Grade A Cash} & {MCI Mumbai} & Bonded Whse Premium Spt\tabularnewline
				{Lead (Metal)} & {LEAD} & {LME Cash} & {MCI Mumbai} & Bonded Whse Premium Spt\tabularnewline
				{Nickel (Metal)} & {NICK} & {LME Cash} &  {MCI Mumbai} & Bonded Whse Premium Spt \tabularnewline
				{Zinc (Metal)} & {ZINC} & {LME 99.995\%} & {MCI Mumbai} & Bonded Whse Premium Spt \tabularnewline
				{Gold (Metal)} & {GOLD} & {Handy\&Hardman Base} & {MCI 99.5\%} & 99.95\% Shanghai\tabularnewline
				{Silver (Metal)} & {SILV} & {Handy\&Hardman (NY)} & {MCI Ahmedabad} & 1\# Shanghai\tabularnewline
				%\hline 
 
					{Corn (Agriculture)} & {CORN} & {No.2 Yellow} & {MCI Nizamabad} & Yellow Pre-Fac Zhengzhou\tabularnewline
					{Soy Oil (Agriculture)} & {SOYB} & {Crude Decatur} & {MCI Indore} & Refined Ex-Fac Dalian\tabularnewline
					{Sugar (Agriculture)} & {SUGA} & {ISA Daily Price} & {MCI Delhi} & Ex-Fac Zhucheng\tabularnewline
					{Wheat (Agriculture)} & {WHEA} & {No.2,Soft Red} & {MCI Delhi Mum} & Bran Ex-Factry Zhengzhou\tabularnewline
					{Cotton (Agriculture)} & {COTT} & {NY Average} & {ICS 101B 22MM Nrt Zn} & SLM CNF China\tabularnewline
					\hline 
				\end{tabular}%
			
\begin{tablenotes}
\item [a] For CRUO and NATG the world price is used, and this for all three markets (due to data availability).
\end{tablenotes}							
\end{threeparttable}
							} % end of resizebox				
		\end{table}
%\end{savenotes}

\begin{table}	
\small
	
	\caption{Portfolio data set. Spot prices of $J=17$ commodities are collected for each of the $K=5$ portfolios. \label{Data_Set_Indices}}
	\vspace{-0.3cm}
	\resizebox{\textwidth}{!}{%	
		\setlength{\tabcolsep}{0.25cm}	
		\begin{tabular}{lllllll}
			\hline 
			\textbf{Index (type)} & \textbf{Label} & \multicolumn{5}{c}{\textbf{Source in Datastream}}\tabularnewline
			\hline 
			\hline 
			\multicolumn{2}{c}{} & \textit{Credit Suisse} & \textit{Dow Jones} & \textit{Merrill Lynch} & \textit{Standard \& Poor's} & \textit{Thomson Reuters}\tabularnewline
			\hline 

				{Agriculture (Global)} & AGRI & {CSCB Spt Ret} & {DJ Capd. Comp.} & {MLCX Spt Indx} & {S\&P GSCI Spt} & {TR Equal Weight CCI}\tabularnewline
				{Energy (Global)} & {ENER} & {CSCB Spt Ret } & {DJ Comm. Indx} & {MLCX Spt Indx} & {S\&P GSCI Spt} & {TR Equal Weight CCI}\tabularnewline
				{Grain (Global)} & {GRAI} & {CSCB Spt Ret} & {DJ Comm. Indx} & {MLCX Spt Indx} & {S\&P GSCI Indx Spt} & {TR Equal Weight CCI}\tabularnewline
				{Ind. Metals (Global)} & {INDM} & {CSCB Spt Ret} & {DJ Comm.} & {MLCX Spt Indx} & {S\&P GSCI Spt} & {TR Equal Weight CCI}\tabularnewline
				%\hline 
					{Crude Oil (Energy)} & {CRUO} & {CSCB Spt Ret} & {DJ Comm.} & {MLCX Spt Indx} & {S\&P GSCI Spt} & {TR/CC CRB Index}\tabularnewline
					{Heating Oil (Energy)} & {HETO} & {CSCB Spt Ret} & {DJ Comm.} & {MLCX Spt Indx} & {S\&P GSCI Spt} & {TR/CC CRB}\tabularnewline
					{Natural gas (Energy)} & {NATG} & {CSCB Spt Ret} & {DJ Comm.} & {MLCX Spt Indx} & {S\&P GSCI Spt} & {TR/CC CRB}\tabularnewline
					%\hline 

						{Aluminum (Metal)} & {ALLU} & {CSCB Spt Ret} & {DJ Comm. Indx} & {MLCX Spt Indx} & {S\&P GSCI Spt} & {TR/CC CRB Index}\tabularnewline
						{Nickel (Metal)} & {NICK} & {CSCB Spt Ret} & {DJ Comm. Indx} & {MLCX Spt Indx} & {S\&P GSCI Spt} & {TR/CC CRB Index}\tabularnewline
						{Gold (Metal)} & {GOLD} & {CSCB Spt Ret} & {DJ Comm. Indx} & {MLCX Spt Indx} & {S\&P GSCI Spt} & {TR/CC CRB Index}\tabularnewline
						{Silver (Metal)} & {SILV} & {CSCB Spt Ret} & {DJ Comm. Indx} & {MLCX Spt Indx} & {S\&P GSCI Spt} & {TR/CC CRB Index}\tabularnewline
						%\hline 
					
							{Corn (Agriculture)} & {CORN} & {CSCB Spt Ret} & {DJ Comm. Indx} & {MLCX Spt Indx} & {S\&P GSCI Spt} & {TR/CC CRB Index}\tabularnewline
							{Wheat (Agriculture)} & {WHEA} & {CSCB Spt Ret} & {DJ Comm. Indx} & {MLCX Spt Indx} & {S\&P GSCI Kansas} & {TR/CC CRB Index}\tabularnewline
							{Soybeans (Agriculture)} & {SOYB} & {CSCB Spt Ret} & {DJ Comm.} & {MLCX Spt Indx} & {S\&P GSCI Spt} & {TR/CC CRB Index}\tabularnewline
							{Sugar (Agriculture)} & {SUGA} & {CSCB \#11 Spt} & {DJ Comm. Indx} & {MLCX Spt Indx} & {S\&P GSCI Spt} & {TR/CC CRB Index}\tabularnewline
							{Coffee (Agriculture)} & {COFF} &
							{CSCB Spt Ret} &
							{DJ Comm. Indx} &
							{MLCX Spt Indx} &
							{S\&P GSCI Spt} &
							{TR/CC CRB Index}
							\tabularnewline
							{Cotton (Agriculture)} & {COTT} & {CSCB Spt Ret} & {DJ Comm. Indx} & {MLCX Spt Indx} & {S\&P GSCI Spt} & {TR/CC CRB Index}\tabularnewline
							\hline 
						\end{tabular}%
					} % end of resize-box
					
				\end{table}

\subsection{Market data set\label{Market_sub}}
\begin{figure}
	\centering
		\includegraphics[scale=0.35]{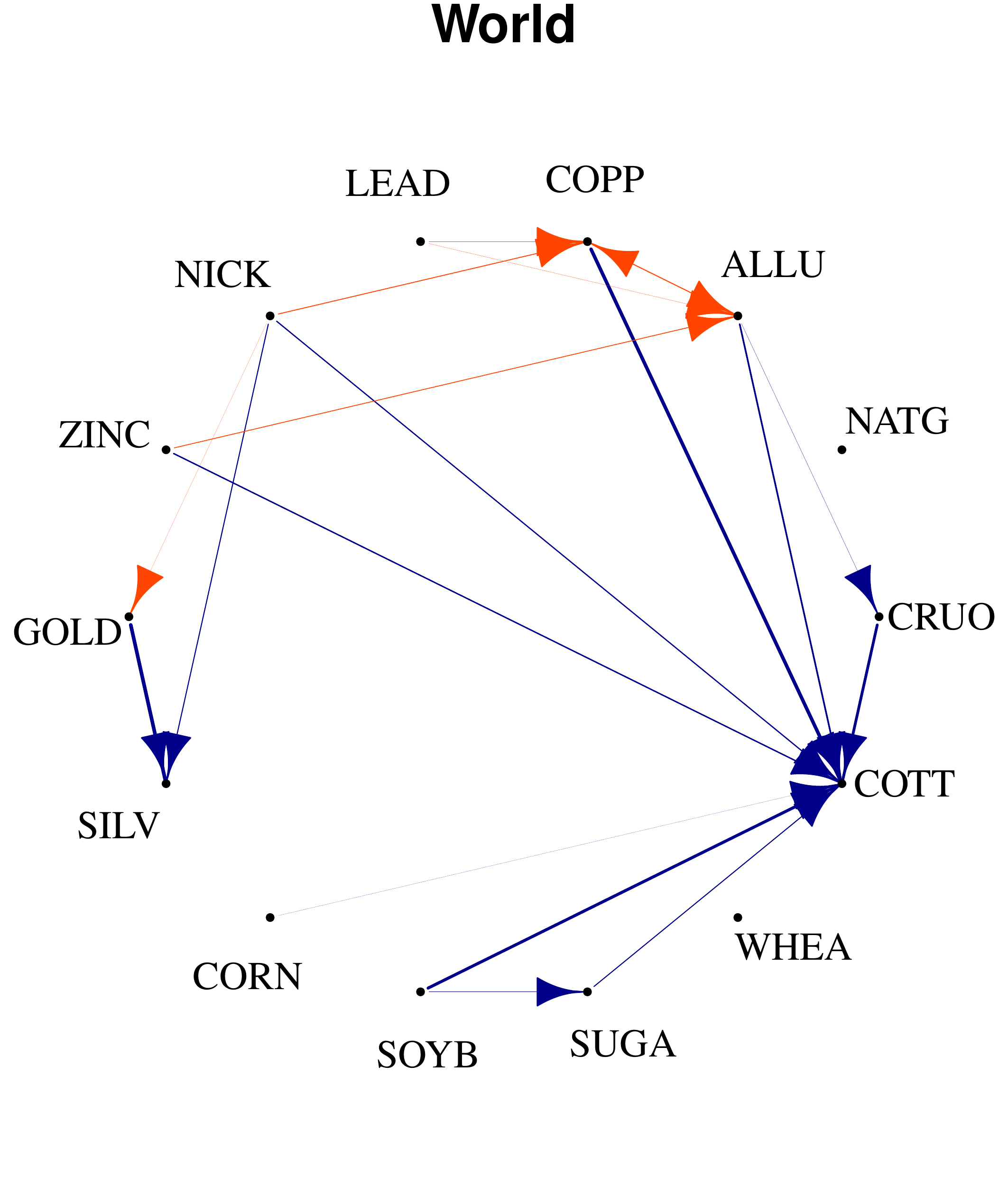}
		\\
		\includegraphics[scale=0.35]{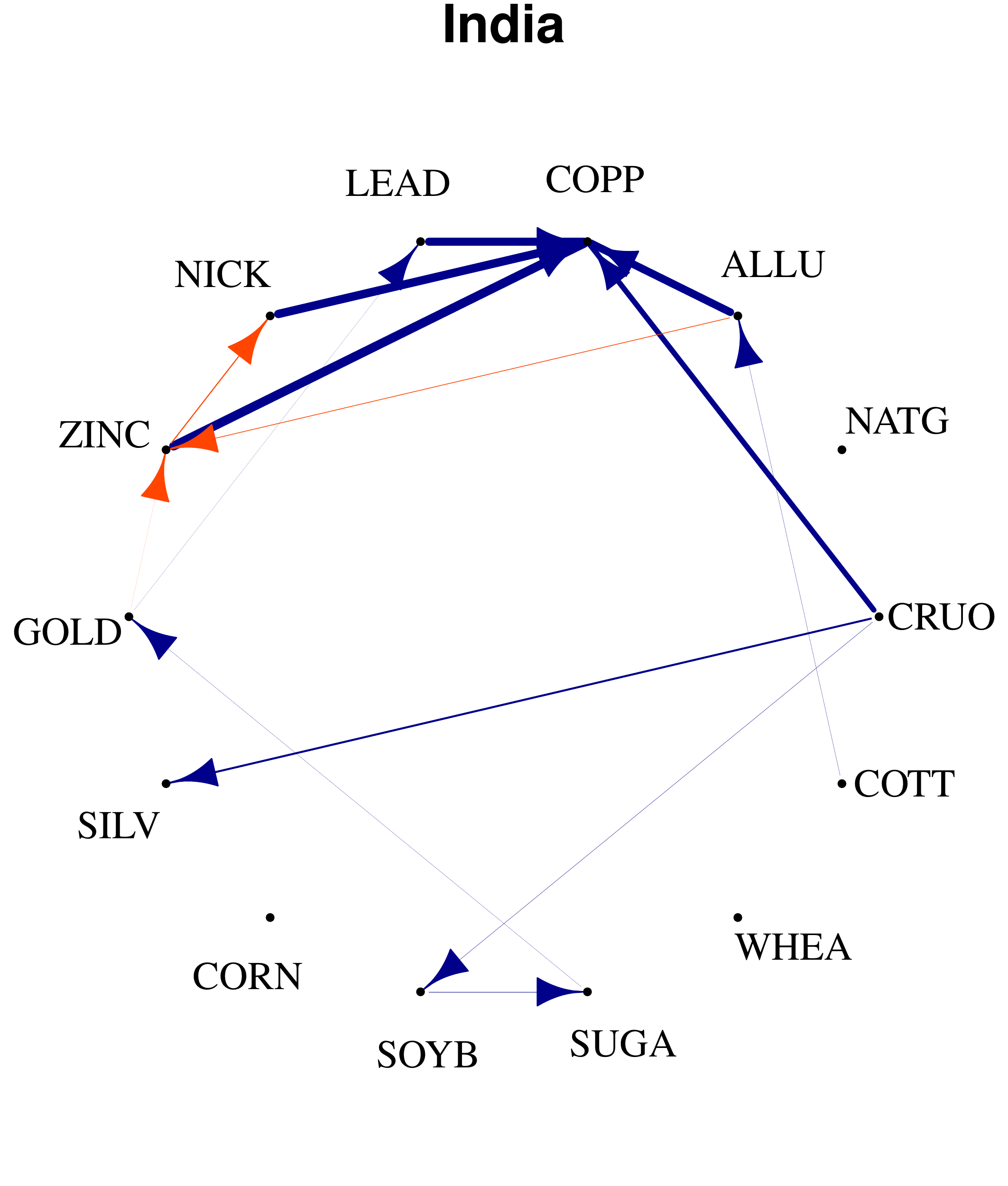}	\hfill
		\includegraphics[scale=0.35]{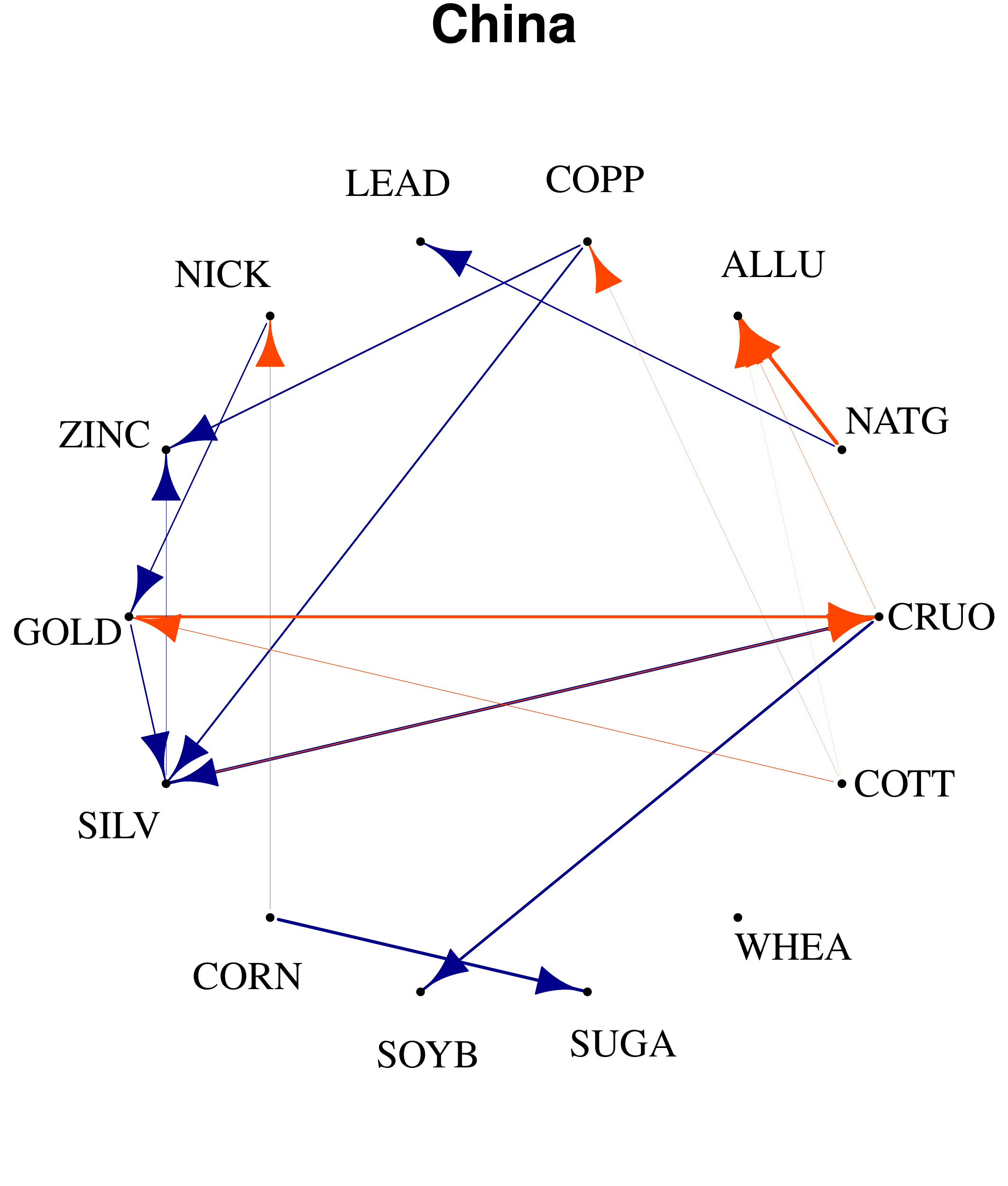}

	\caption{Market data set. Commodity effect networks: a directed edge is drawn from one commodity to another if the associated commodity price effect is estimated non-zero. The edge width represents the magnitude of the effect. Positive effects are shown in blue (dark gray) and negative effects in red (light gray). \label{AreaColorWeighted}}
\end{figure}

We use daily spot returns ranging from November 1st, 2013 to November 2nd, 2015, hence, $T=522$ time observations of three commodity types (energy, metal and agricultural), see Table \ref{Data_Set_Countries}. The price effect networks for the three markets are presented in Figure \ref{AreaColorWeighted}. For instance in the \textit{World network},  an edge is drawn from gold to silver since the effect of the first on the second is estimated as non-zero.

% Shared Sparsity Patterns
\paragraph{Market comparison}
Table \ref{Areas_shared_effects} reports the proportions of shared non-zero  effects among the three markets.
The networks show limited evidence of shared commodity effects in the three markets. 
India shares only 33\% of its effects both with World and China. Besides, only 17\% of the effects observed in China are also present in the World network. This little evidence of shared effects between markets is in line with the existing literature. \cite{Bukenya05} underlines the existence of few similarities between local and world markets for agricultural and metal commodity spot price effects. \cite{Achvarina06} derive similar conclusions for the Chinese and World aluminum markets.  

\begin{table}
		\caption{ Market data set. Proportions of shared non-zero effects among markets: each cell indicates the proportion of effects for market $i$ (row) that are also present for market $j$ (column).
\label{Areas_shared_effects}}
		\medskip
		\centering
		\begin{tabular}{c|ccc}
			\hline
			& World & India & China \\ 
			\hline
			World & 1.00 & 0.26 & 0.17 \\ 
			India & 0.33 & 1.00 & 0.33 \\ 
			China & 0.17 & 0.25 & 1.00 \\ 
			\hline
		\end{tabular}
\end{table}

\begin{figure}[h]
\centering
\resizebox{.87\textwidth}{!}{%	
\begin{subfigure}{.33\textwidth}
	\centering
	%\resizebox{14cm}{!}{%
	\includegraphics[width=\linewidth]{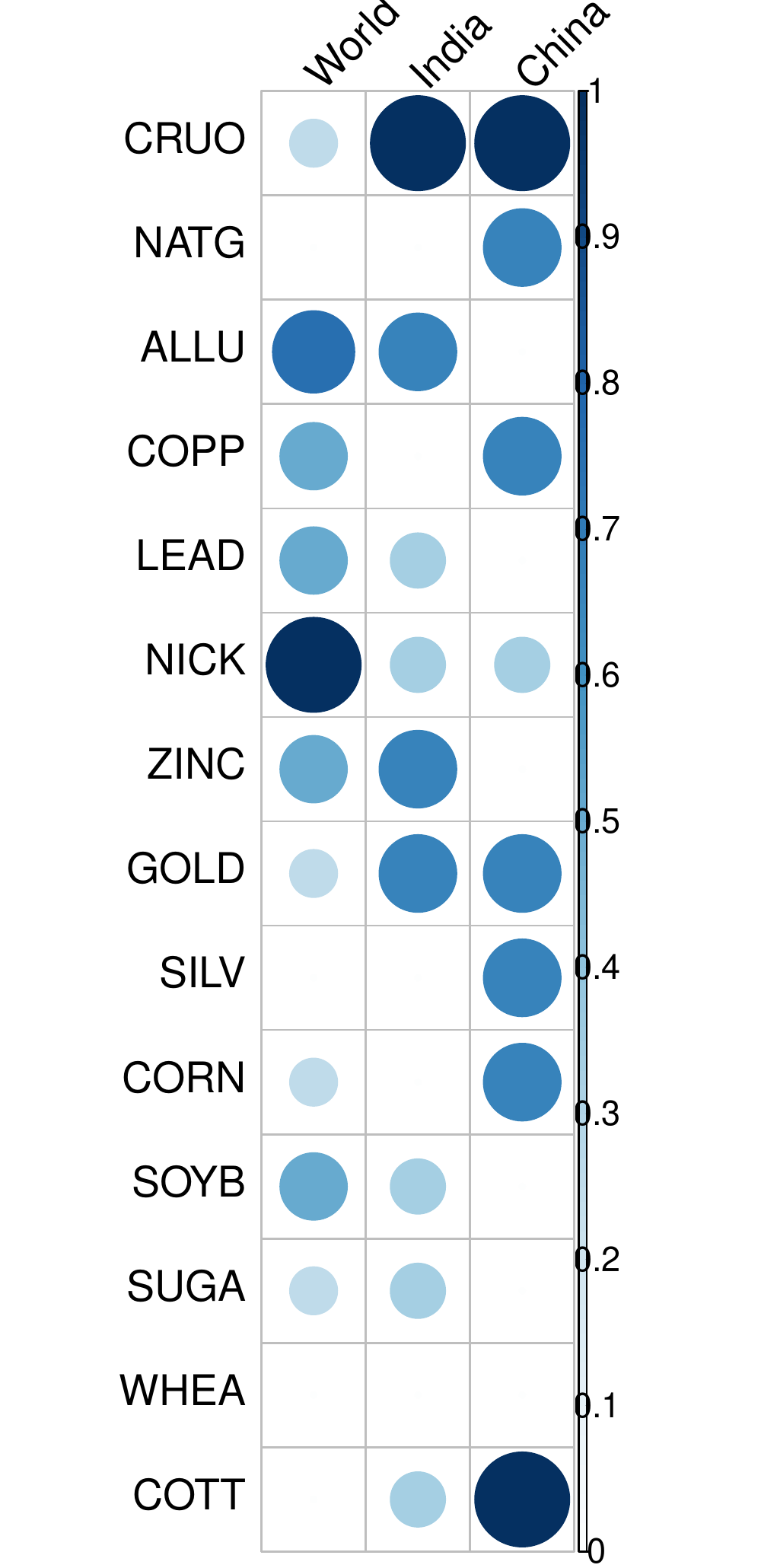}
	\caption{ Out-going connectedness}
	\label{areas_out}
\end{subfigure}%
\begin{subfigure}{.33\textwidth}
	\centering
	\includegraphics[width=\linewidth]{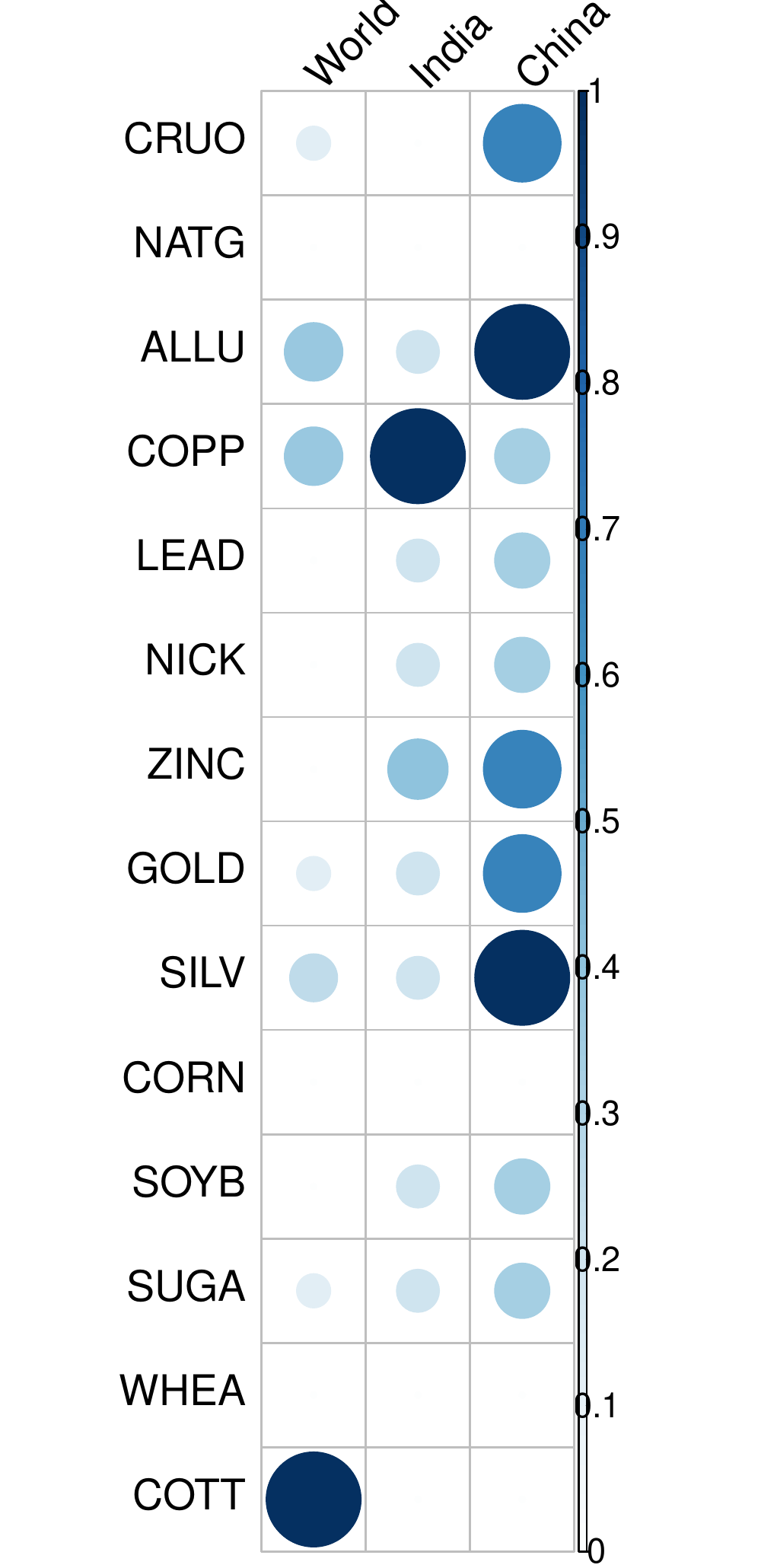}
	\caption{In-going connectedness}
	\label{areas_in}
\end{subfigure}
\begin{subfigure}{.33\textwidth}
	\centering
	\includegraphics[width=\linewidth]{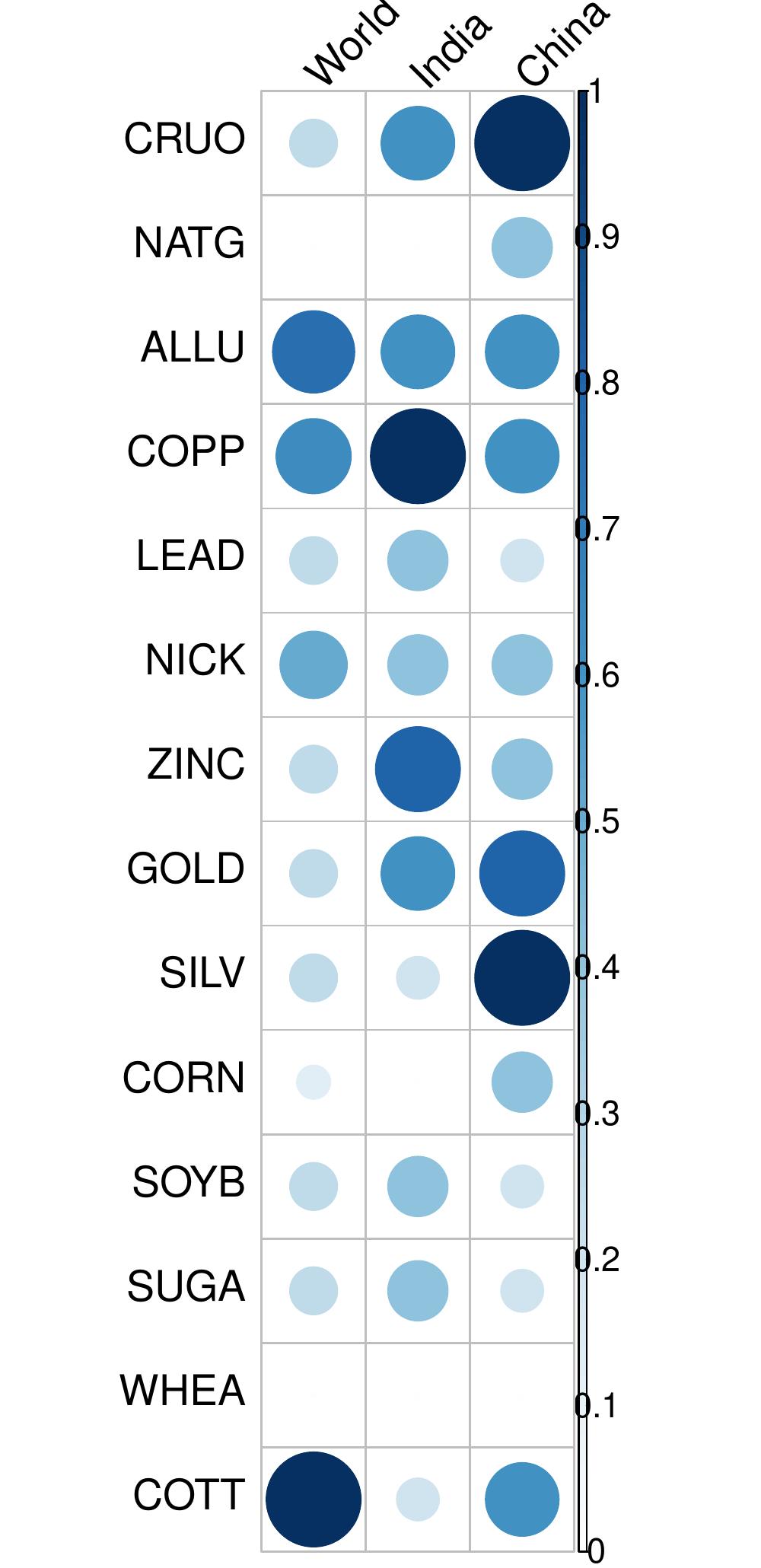}
	\caption{Total connectedness}
	\label{areas_tot}
\end{subfigure}%
}% end of resizebox

\caption{Market data set. Measures of (a) out-going, (b) in-going and (c) total connectedness for each commodity (rows) in each market (columns). The size (and the color) of the circle reflects the magnitude of the connectedness: the larger (and darker), the more connected the commodity.}
\label{areas_connect}
\end{figure}

% Connectedness 
\paragraph{Commodity connectedness}
Figure \ref{areas_connect} pictures the measures of connectedness (cfr. Subsection \ref{Network_interpretation}) for each market. 
The most connected commodities are mainly found among metals for all three markets, either by looking at out-going, in-going or total connectedness. This finding is coherent with our expectations on the Indian and Chinese markets, since both Asian economies heavily depend on their metallurgic sectors. The Shanghai Metal Exchange is one of the largest commodity exchanges in the world for metals \citep{Klotz14}: China represents the biggest producer and consumer of different metals and minerals worldwide \citep{Pitfield10}. Similar considerations are found for India \citep{Jain13, Pitfield10}. 
Besides, crude oil presents high out-going connectedness in the Indian and Chinese markets, signaling that the crude oil price is a leading factor of commodity price dynamics in the two Asian economies.

With respect to agricultural commodities, cotton (COTT) is the most connected commodity in terms of total connectedness. In China (and to a lesser extent India) this is due to the presence of high out-going connectedness (there is no in-going connectedness in both markets), whereas in World there is a higher prevalence of in-going connectedness. 
The different behavior of cotton markets could be due to different export strategies. Indeed, Indian and Chinese cotton mostly satisfy the local demand (respectively, only 14\% and 0.2\% of the total cotton production in 2014 was exported).
On the contrary, US cotton, which we take as the world benchmark, is largely exported to international markets (69\% of the US total cotton production in 2014 was exported).
Therefore, it largely depends on the world demand conditions and is more responsive to other commodity behavior.

\paragraph{Effects across commodity types}

% Within
Table \ref{Areas_cross_effects} reports on the main diagonal the proportions of non-zero effects within each commodity type for each market. 
For instance, out of the 20 possible effects from one agricultural commodity towards another agricultural commodity, only four of them are estimated as non-zero in the World network, hence there are 20\% of non-zero within agriculture effects.
In the World network non-zero within effects involve both metal (21\%) and agricultural commodities (20\%), whereas in India and China there is a higher prevalence of non-zero effects within metal commodities (19\% and 12\% respectively) than within agricultural commodities (both 5\%). Hereby, it is confirmed the relevance of the metal sector in the two Asian economies. In all three markets, there are no effects within energy commodities.

% Spillover
Table \ref{Areas_cross_effects} reports on the off-diagonal entries the proportions of spillover effects among commodity types for each market. 
For instance, out of the 10 possible effects from an energy commodity towards an agricultural commodity, only one of them is estimated as non-zero in the World network, hence 10\% of the spillovers effects are estimated as non-zero.
In all three networks we observe strong spillover effects in commodity returns from energy towards agriculture (10\% in each market), confirming 
the studies of \cite{Tyner10, Serra13} and references therein.
Furthermore, in the Indian and Chinese network there are relevant spillover effects from energy towards metal (14\% and 29\% respectively). This finding is in line with previous studies where energy prices are found to be highly influential towards industrial metal commodities \citep{Akram09, Klotz14} and precious metals \citep{Sari10}. 
Besides, the Indian and Chinese markets present unidirectional spillover effects from agriculture towards metal (6\% and 11\% respectively): this asymmetric structure of the spillover effects is outspoken and might reflect an additional effect from energy towards metal via agricultural commodities.

This asymmetric spillover structure is not present in the World network. The high proportion of spillover effects from metal to agriculture (11\%) is mainly driven by the high in-going connectedness of cotton (cfr. Table \ref{areas_connect}).  
However, cotton represents a special case. Indeed, when cotton is traded no actual delivery happens and the cotton price is the result of an average of offer quotations \citep{Bukenya05}. This might result in distorted dynamics involving cotton.

\begin{table}

	\caption{Market data set. For each market, the proportion of non-zero effects within each commodity type (diagonal elements) and spillover effects across commodity types (off-diagonal elements) are given. \label{Areas_cross_effects}}
	\medskip
	\centering
	\small
	\resizebox{0.46\textwidth}{!}{%

				\begin{tabular}{c|c|ccc}
				\hline 
				\multicolumn{5}{c}{\textbf{World}}\tabularnewline
				\hline 
				\hline 
				& \multicolumn{4}{c}{\small \textit{To}}\tabularnewline
				\hline 
				\multirow{4}{*}{\begin{turn}{90}
						{\small\textit{From}}
					\end{turn}}
					&  & Energy & Metal & Agriculture\tabularnewline
					\cline{2-5}
					& Energy & 0.00 & 0.00 & 0.10 \\ 
					& Metal & 0.07 & 0.21 & 0.11 \\ 
					& Agriculture & 0.00 & 0.00 & 0.20 \\
					
				\end{tabular} %
			}

			\bigskip

	\resizebox{\textwidth}{!}{%	
			\begin{tabular}{c|c|ccc}
				\hline 
				\multicolumn{5}{c}{\textbf{India}}\tabularnewline
				\hline 
				\hline 
				& \multicolumn{4}{c}{\small \textit{To}}\tabularnewline
				\hline 
				\multirow{4}{*}{\begin{turn}{90}
						{\small\textit{From}}
					\end{turn}}
					&  & Energy & Metal & Agriculture\tabularnewline
					\cline{2-5}
					& Energy & 0.00 & 0.14 & 0.10 \\ 
					& Metal & 0.00 & 0.19 & 0.00 \\ 
					& Agriculture & 0.00 & 0.06 & 0.05 \\
					
				\end{tabular} %
					
				\hspace{1.5cm}
					
					\begin{tabular}{c|c|ccc}
						\hline 
						\multicolumn{5}{c}{\textbf{China}}\tabularnewline
						\hline 
						\hline 
						& \multicolumn{4}{c}{\small \textit{To}}\tabularnewline
						\hline 
						\multirow{4}{*}{\begin{turn}{90}
								{\small\textit{From}}
							\end{turn}}
							&  & Energy & Metal & Agriculture\tabularnewline
							\cline{2-5}
							& Energy & 0.00 & 0.29 & 0.10 \\ 
							& Metal & 0.14 & 0.12 & 0.00 \\ 
							& Agriculture & 0.00 & 0.11 & 0.05 \\ 
						\end{tabular} %
			} % end of resizebox
				
			\end{table}

\subsection{Portfolio data set}
\begin{figure}
		\centering
		\includegraphics[scale=0.26]{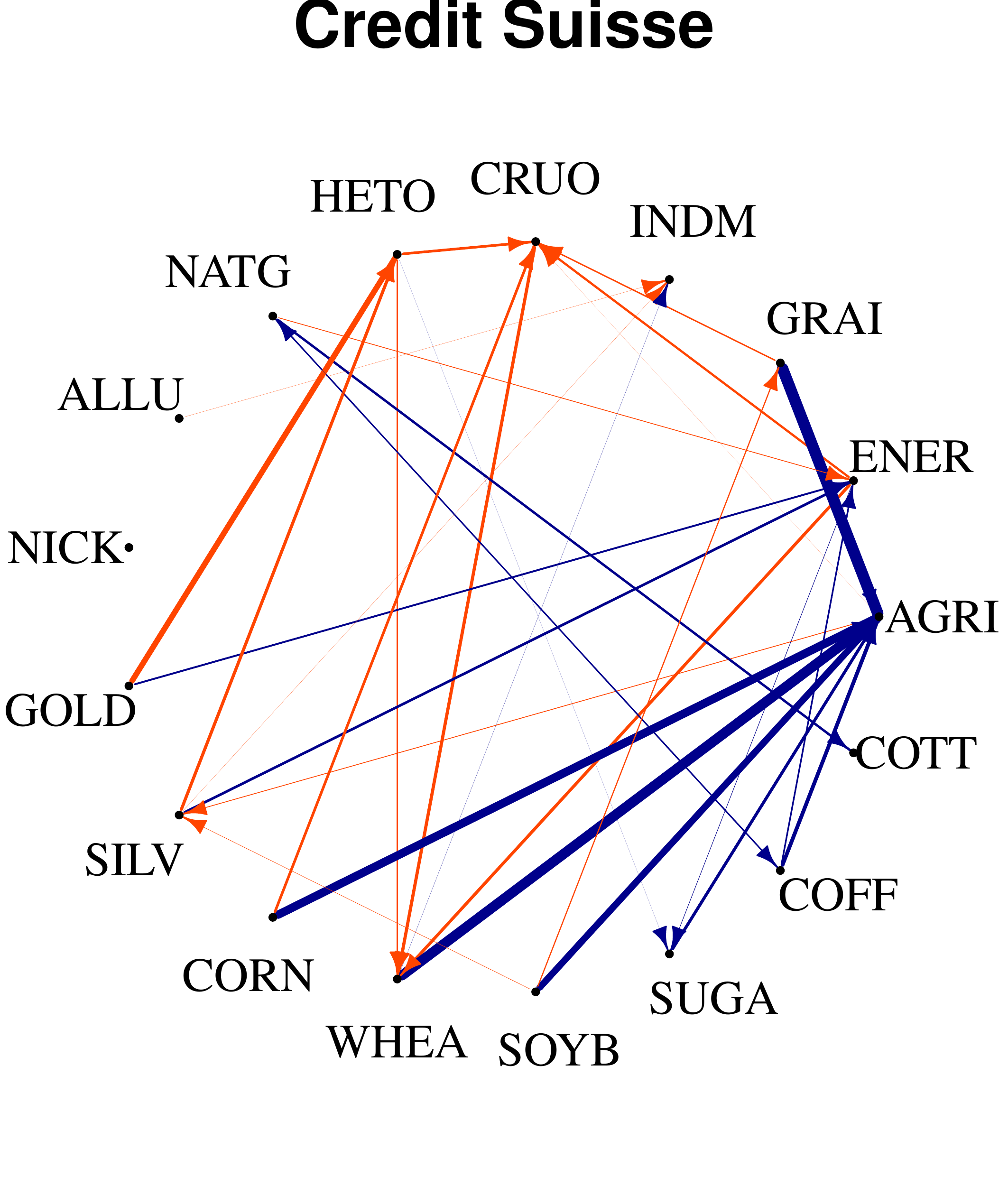} \hspace{1.5cm}
		\includegraphics[scale=0.26]{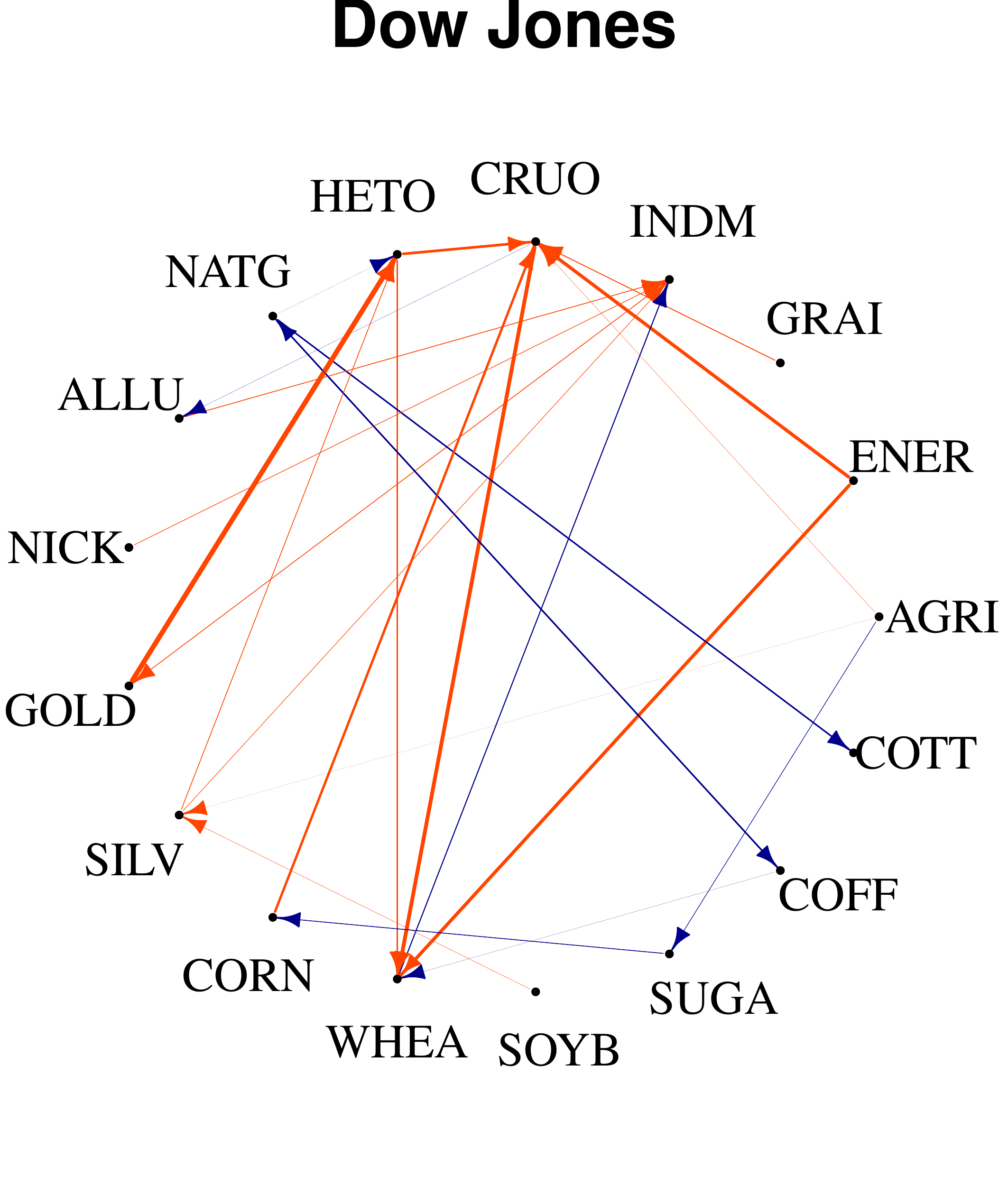}
		\\
		\includegraphics[scale=0.26]{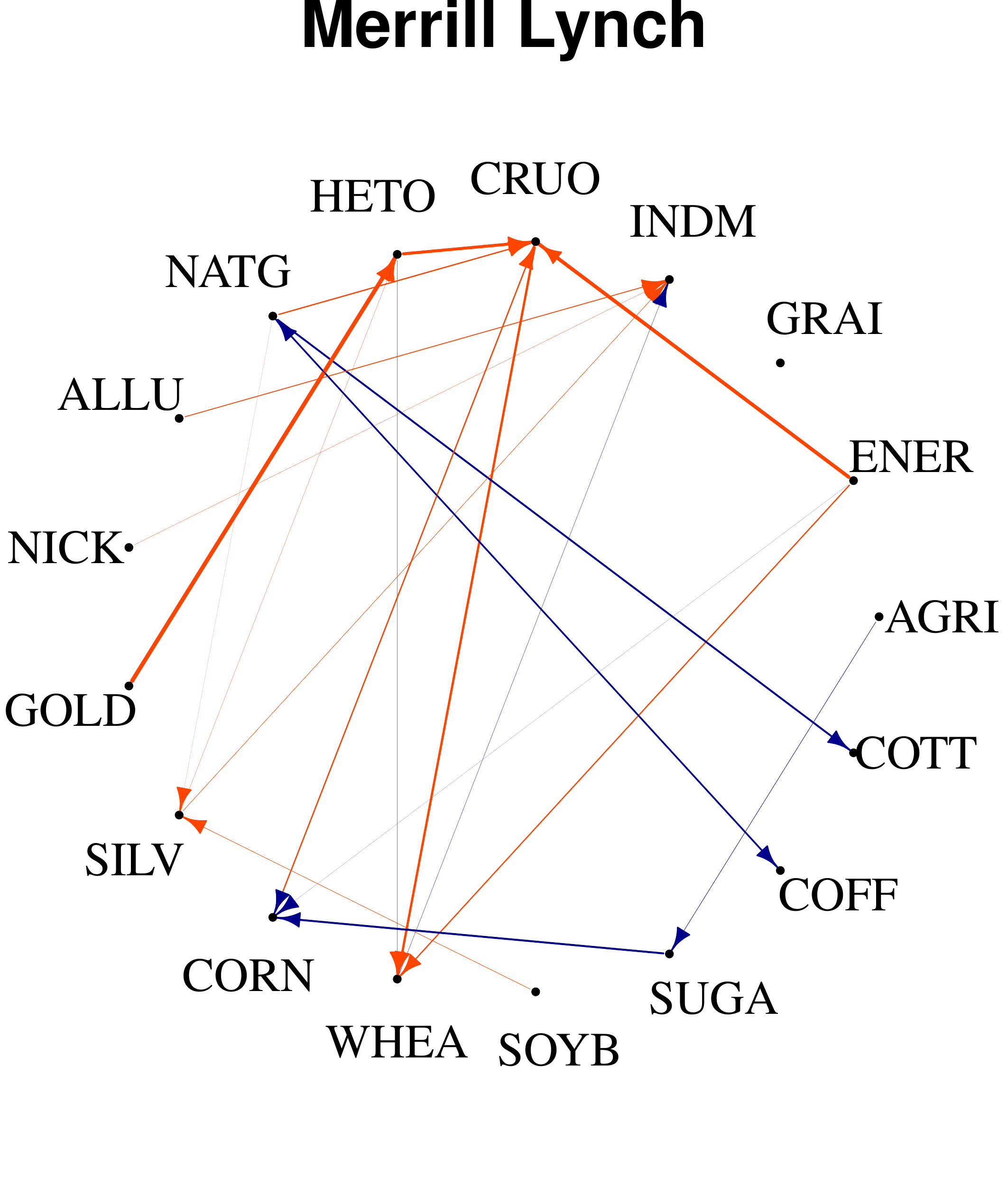} \hspace{1.5cm}
		\includegraphics[scale=0.26]{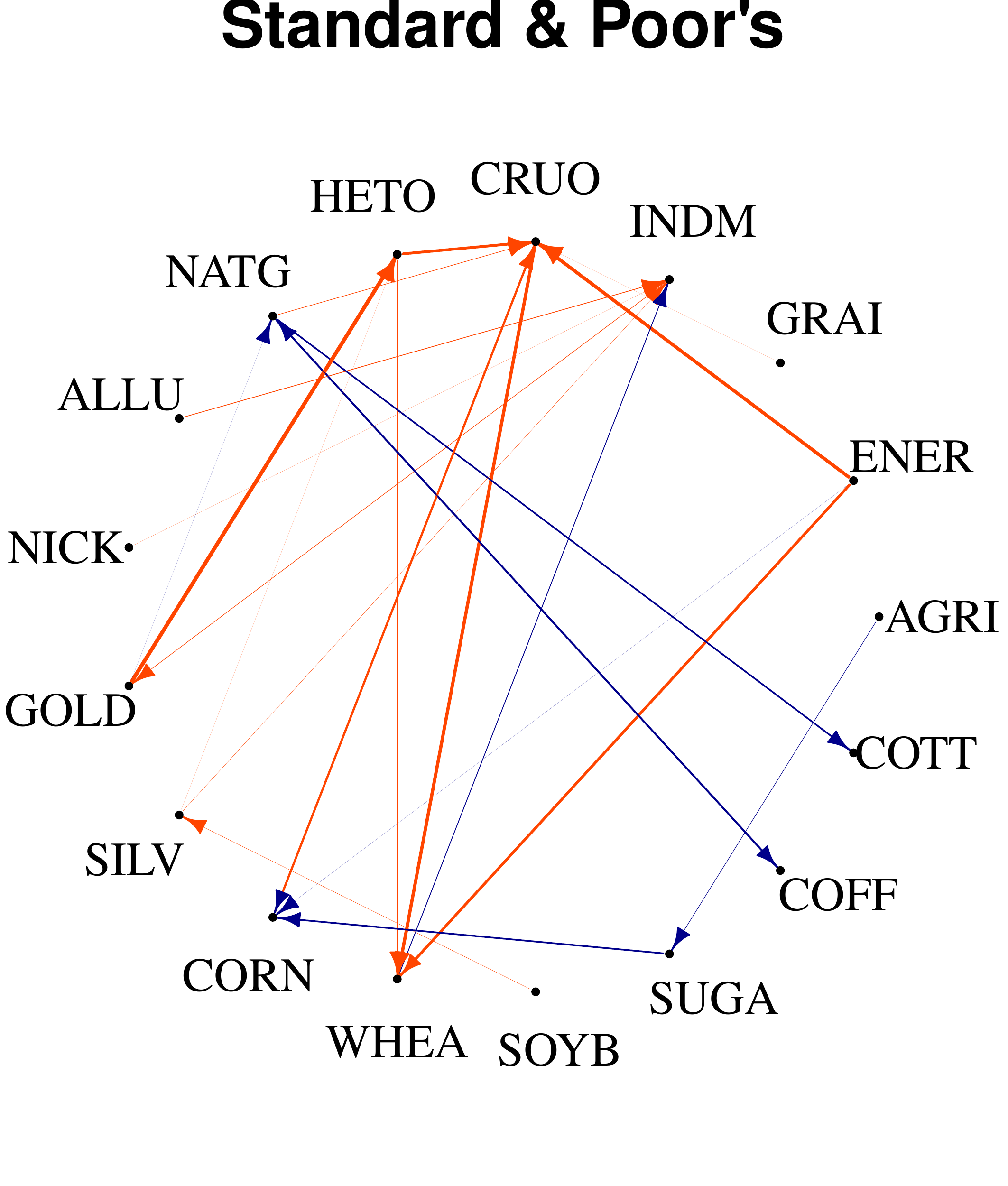} 
		\\
		\includegraphics[scale=0.26]{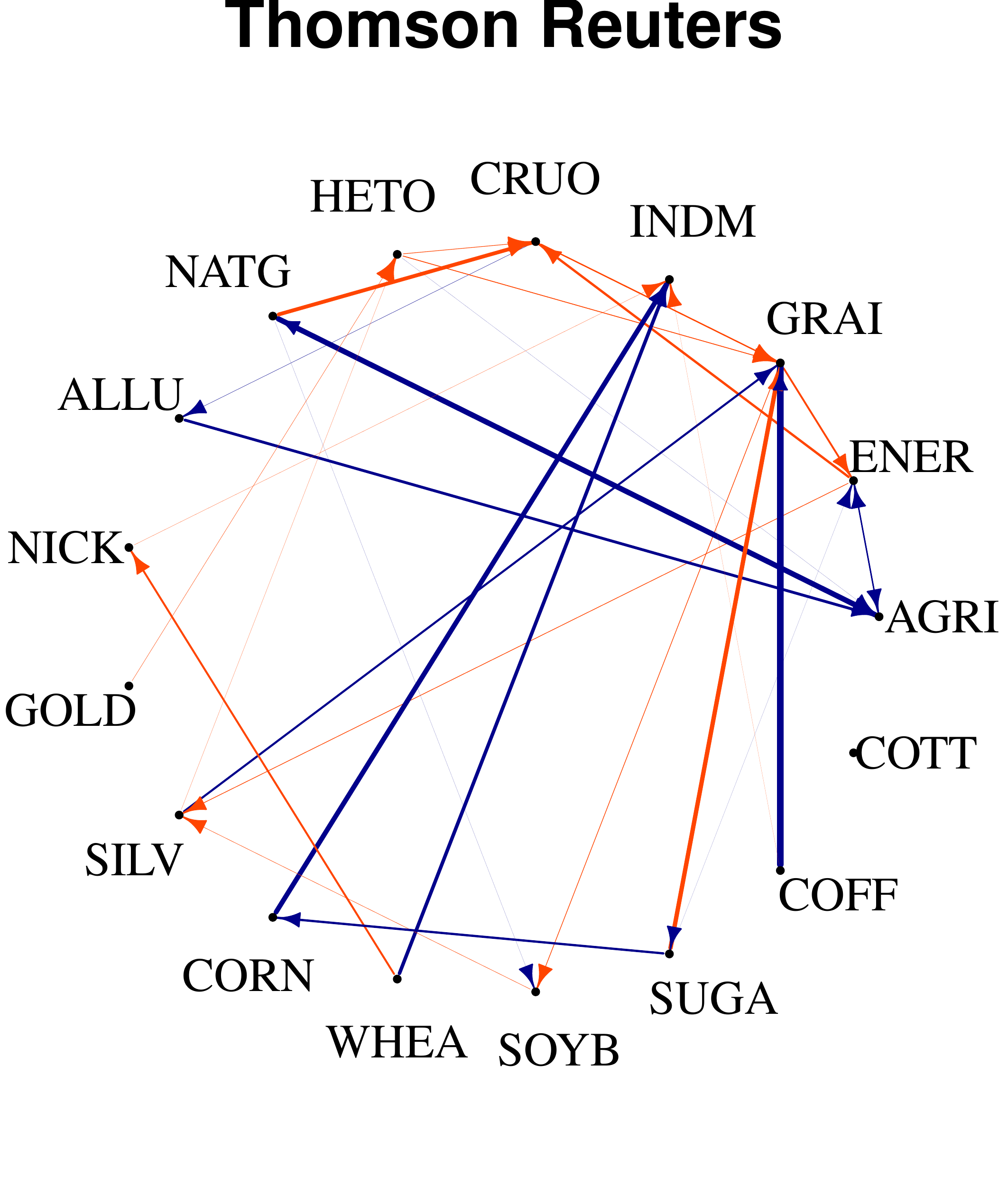}
		
		%\vspace{-0.8cm}
		\caption{Portfolio data set. Commodity effect networks: a directed edge is drawn from one commodity to another if the associated price effect is estimated non-zero. The edge width represents the magnitude of the effect. Positive effects are shown in blue (dark gray) and negative effects in red (light gray). \label{Comm_Indexes_Network}}
\end{figure}

We use daily spot return indexes ranging from November 1st, 2013 to November 2nd, 2015, hence, $T=522$ time observations of four different commodity types (global index, energy, metal, agriculture), see Table \ref{Data_Set_Indices}. The price effect networks for the five portfolios are presented in Figure \ref{Comm_Indexes_Network}. 

% Shared Sprsity patterns
\paragraph{Portfolio comparison} 
Table \ref{Indexes_shared_effects} reports the proportions of shared non-zero  effects among the five portfolios.
The networks show much more evidence of shared effects in the portfolios (on average 56\%) compared to the market data set (on average 25\%). 
For instance, 89\% of the effects observed in Standard \& Poor's are also present in Merrill Lynch. Moreover, the direction, the sign and the magnitude of the effects are comparable in the different portfolios. 
Although each commodity index is differently built, they present similar features and guarantee comparable performances \citep{Anson06}. The only exception is represented by Thomson Reuters, with smaller proportions of shared effects with the other portfolios\footnote{For instance, only 24\% of the effects observed in Thomson Reuters are also observed in Credit Suisse.}. Moreover, Thomson Reuters has non-zero commodity effects that differ in terms of sign, direction and magnitude. Since the index weighting scheme of Thomson Reuters is not publicly available, we cannot examine this anomaly. If we exclude Thomson Reuters from the analysis, then the four remaining portfolios have on average 71\% of shared effects. 

\begin{table}
	\caption{Portfolio data set. Proportions of shared non-zero effects among portfolios: each cell indicates the proportion of effects for portfolio $i$ (row) that are also present for portfolio $j$ (column). \label{Indexes_shared_effects}}
	\medskip
\resizebox{0.9\textwidth}{!}{\begin{minipage}{\textwidth}
\centering
	\begin{tabular}{l|ccccc}
		\hline
		& Credit Suisse & Dow Jones & Merrill Lynch & Standard \& Poor's & Thomson Reuters \\ 
		\hline
		Credit Suisse & 1.00 & 0.56 & 0.47 & 0.50 & 0.23 \\ 
		Dow Jones & 0.66 & 1.00 & 0.76 & 0.83 & 0.38 \\ 
		Merrill Lynch & 0.61 & 0.85 & 1.00 & 0.96 & 0.42 \\ 
		Standard \& Poor's & 0.61 & 0.86 & 0.89 & 1.00 & 0.39 \\ 
		Thomson Reuters & 0.24 & 0.33 & 0.33 & 0.33 & 1.00 \\ 
		\hline
	\end{tabular}
	\end{minipage} }
\end{table}

\begin{figure}[t]
	\centering
	\resizebox{0.87\textwidth}{!}{%	
	\begin{subfigure}{.33\textwidth}
		\centering
		\includegraphics[width=\linewidth]{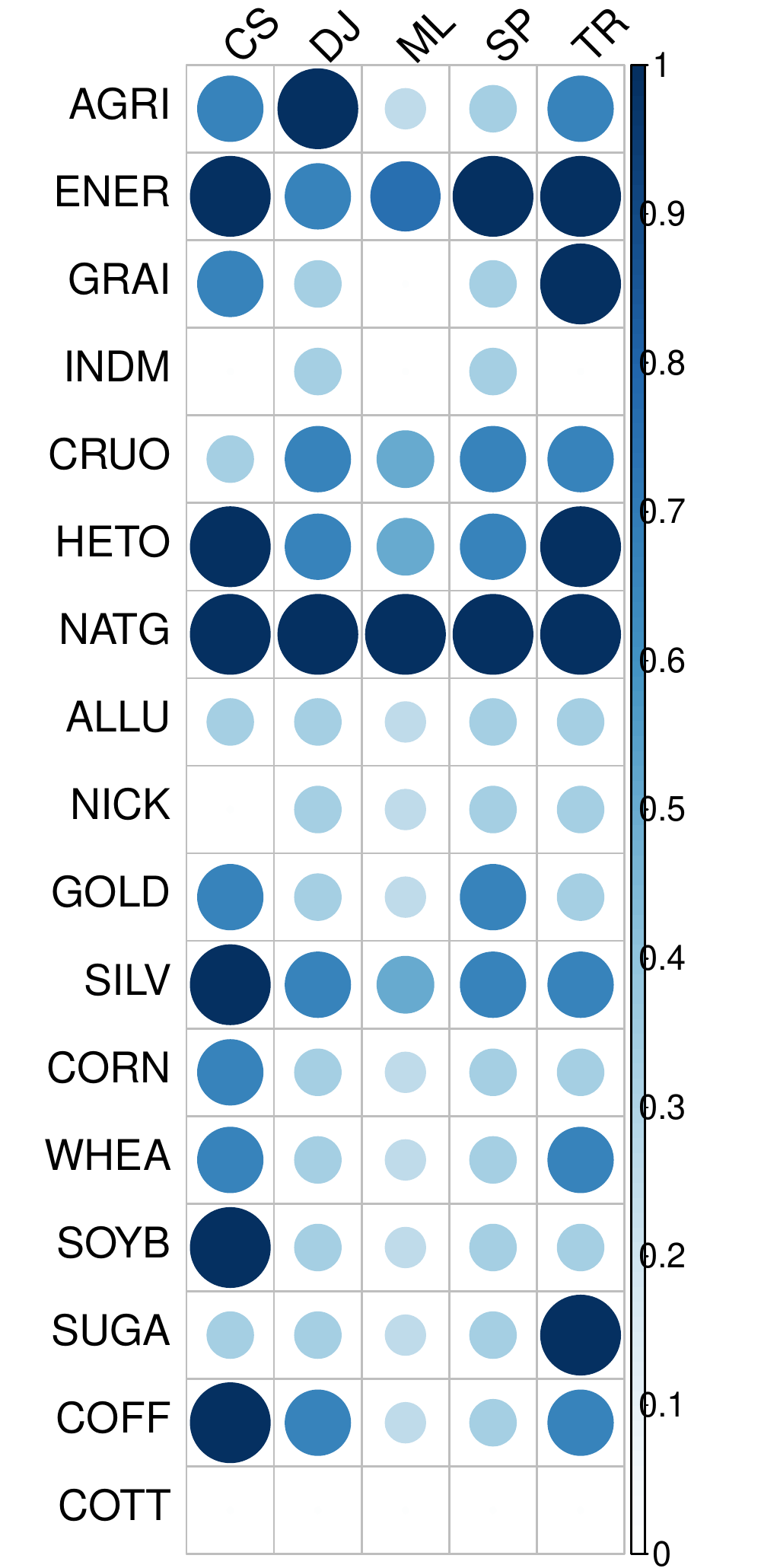}
		\caption{Out-going connectedness}
		\label{indexes_out}
	\end{subfigure}%
	\begin{subfigure}{.33\textwidth}
		\centering
		\includegraphics[width=\linewidth]{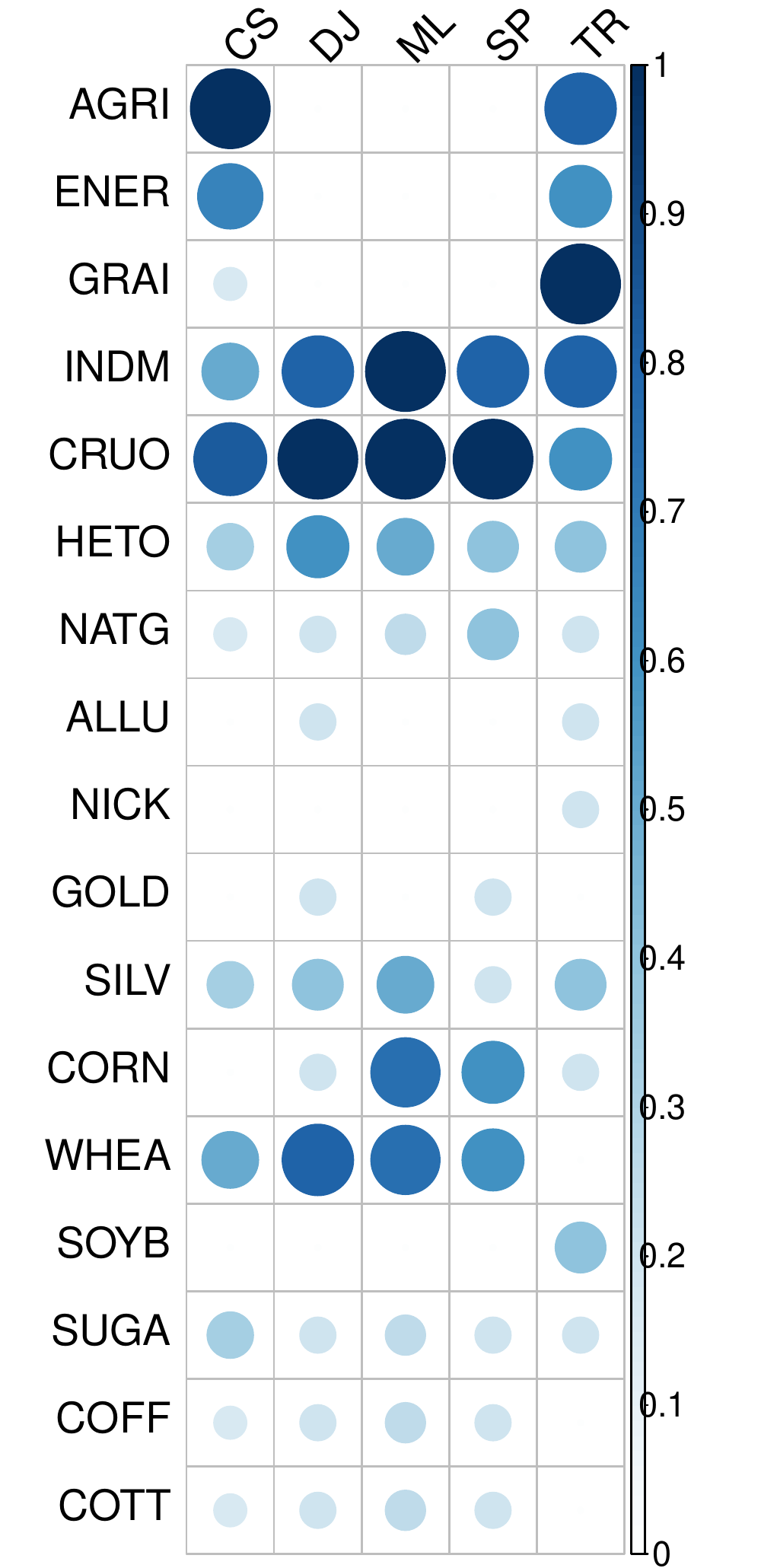}
		\caption{In-going connectedness}
		\label{indexes_in}
	\end{subfigure}
	\begin{subfigure}{.33\textwidth}
		\centering
		\includegraphics[width=\linewidth]{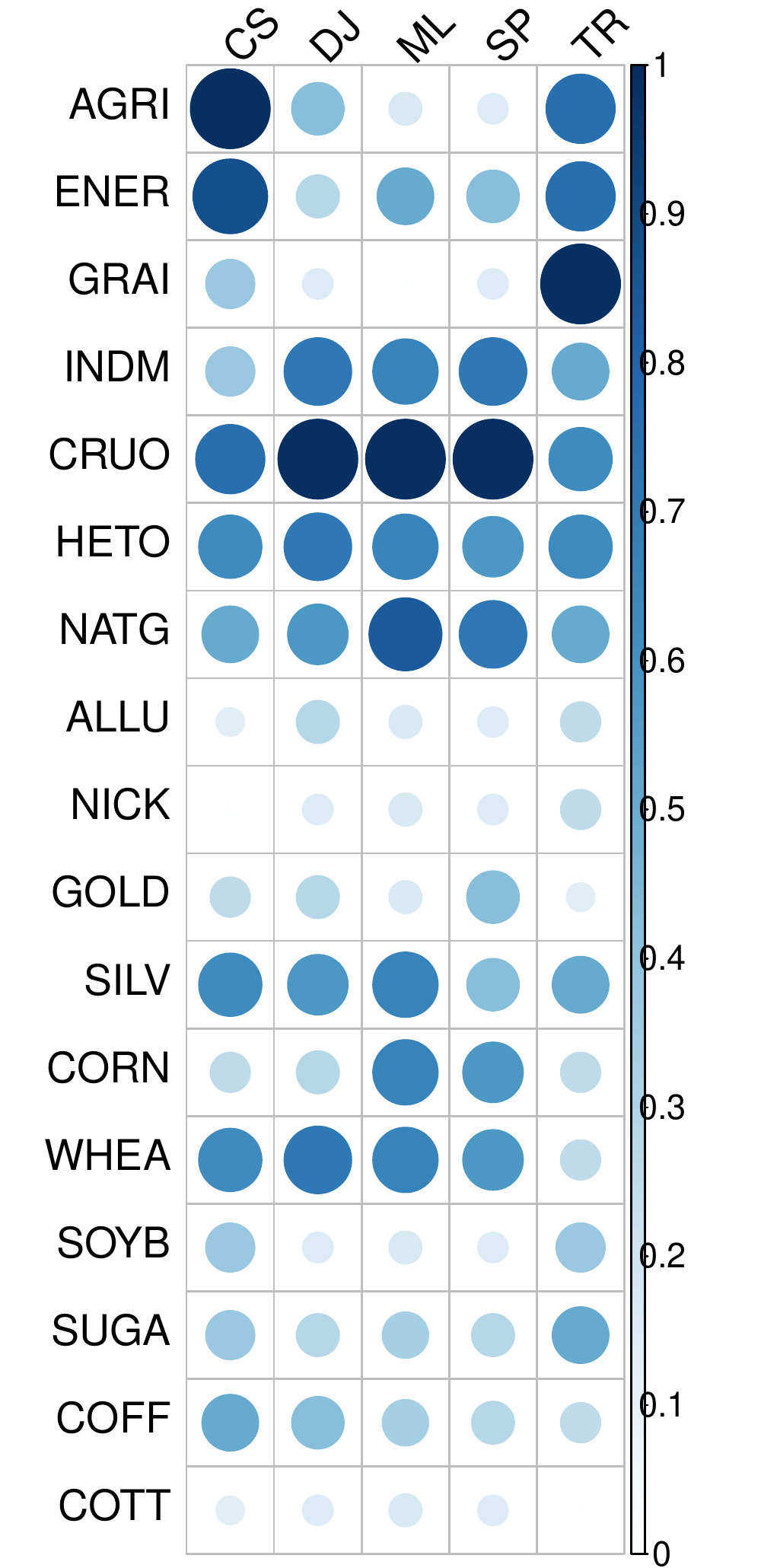}
		\caption{Total connectedness}
		\label{indexes_tot}
	\end{subfigure}%
	} % end of resizebox
	
	\caption{Portfolio data set. Measures of (a) out-going, (b) in-going and (c) total connectedness for each commodity (rows) in each portfolio (columns). The size (and the color) of the circle reflects the measure of connectedness: the larger (and darker), the more connected the commodity.}
	\label{indexes_connect}
\end{figure}

% Connectedness
\paragraph{Commodity connectedness} 
Figure \ref{indexes_connect} pictures the three measures of connectedness for each commodity index in each portfolio. 
In the majority of the cases, the most connected commodities, in terms of total connectedness, are either found among global or energy indexes.
Global indexes have in general higher out-going than in-going connectedness: the most connected indexes are the energetic (ENER) and agricultural ones (AGRI and GRAI). 
Overall, these agricultural and the energy global indexes play a relevant role in our network analysis. Moreover, the high out-going connectedness of global indexes suggests that they are driving forces among the set of commodities in our analysis.
Energy indexes show high total connectedness, which is either due to high out-going connectedness, as for natural gas (NATG), or to high in-going connectedness, as for crude oil (CRUO).
Agriculture commodities show a moderate out-going connectedness, in particular soybeans (SOYB), sugar (SUGA) and coffee (COFF), and moderate in-going connectedness, in particular wheat (WHEA) and corn (CORN).
Metal commodities show overall little total connectedness, with the only exception of silver (SILV).

\paragraph{Effects across commodity types}
% Within
Table \ref{Indexes_cross_effects} reports on the main diagonal the proportions of non-zero effects within each commodity type for each portfolio. 
The largest proportion of non-zero effects is estimated within energy commodities in each portfolio, thus indicating that a large part of the total connectedness of energy commodities is explained by effects among crude oil, natural gas and heating oil.\footnote{This finding might seem contradictory to the market data application where no within energy effects were found. Here, however, all within energy effects involve heating oil which was - due to data availability - not included in the market data set. In line with the market application, we found no effects between natural gas and crude oil.}
Apart from Credit Suisse, all portfolios show non-zero effects within agricultural commodities. In particular, there is an outspoken positive effect from sugar towards corn: a sugar price increase is reflected in a corn price increase. This relation is not surprising since both sugar and corn are used in ethanol production: they are not substitute goods (since the production process of ethanol is different when using sugar or corn), but they both respond to the same price dynamics, thereby explaining the positive sign of the effect.

% Spillover
Table \ref{Indexes_cross_effects} reports on the off-diagonal entries the proportions of spillover effects among commodity types for each portfolio.
In all portfolios spillover effects are estimated from global towards agriculture: for instance, 12\% of the spillover effects in this direction are estimated non-zero in Standard \& Poor's. The majority of these spillover effects derive from the global index for energy (cfr. Figure \ref{Comm_Indexes_Network}).
Moreover, spillover effects are also present from energy towards agriculture in all portfolios: 28\%  of the spillover effects in this direction are estimated non-zero in Credit Suisse, Merrill Lynch and Standard \& Poor's. 
Our analysis not only confirms the existence of spillovers from energy towards agricultural, as in \cite{Chen10, Nazlioglu12}, but also finds evidence of spillover effects in the opposite direction, as in  \cite{Rezitis15}. Indeed, 11\% of the spillover effects from agriculture towards energy are estimated non-zero in all portfolios apart from Thomson Reuters.

\begin{table}[t]

	\caption{Portfolio data set. For each portfolio, the proportion of non-zero effects within each commodity type (diagonal elements) and spillover effects across commodity types (off-diagonal elements) are given.\label{Indexes_cross_effects}}
	\resizebox{\textwidth}{!}{%
		\begin{tabular}{c|c|cccc}
			\hline 
			\multicolumn{6}{c}{\textbf{Credit Suisse}}\tabularnewline
			\hline 
			\hline 
			& \multicolumn{5}{c}{\textit{\small{}To}}\tabularnewline
			\hline 
			\multirow{5}{*}{\begin{turn}{90}
					{\small\textit{From}}
				\end{turn}}
			&  & Global & Energy & Metal & Agriculture\tabularnewline
			\cline{2-6} 
			& Global & 0.08 & 0.25 & 0.06 & 0.08 \\ 
			& Energy & 0.08 & 0.17 & 0.00 & 0.28 \\ 
			& Metal & 0.25 & 0.17 & 0.00 & 0.00 \\ 
			& Agriculture & 0.33 & 0.11 & 0.04 & 0.00 \\ 
		\end{tabular}
	\ \ \ \ \ \ %
		\begin{tabular}{c|c|cccc}
			\hline 
			\multicolumn{6}{c}{{\textbf{Dow Jones}}}\tabularnewline
			\hline 
			\hline 
			& \multicolumn{5}{c}{\textit{\small{}To}}\tabularnewline
			\hline 
			\multirow{5}{*}{\begin{turn}{90}
					{\small\textit{From}}
				\end{turn}}
			&  & Global & Energy & Metal & Agriculture\tabularnewline
			\cline{2-6} 
			& Global & 0.00 & 0.25 & 0.12 & 0.08 \\ 
			& Energy & 0.00 & 0.33 & 0.08 & 0.22 \\ 
			& Metal & 0.19 & 0.17 & 0.00 & 0.00 \\ 
			& Agriculture & 0.04 & 0.11 & 0.04 & 0.07 \\
		\end{tabular}
		} 
	\bigskip	
		
	\resizebox{\textwidth}{!}{%
		\begin{tabular}{c|c|cccc}
			\hline 
			\multicolumn{6}{c}{\textbf{Merrill Lynch}}\tabularnewline
			\hline 
			\hline 
			& \multicolumn{5}{c}{\textit{\small{}To}}\tabularnewline
			\hline 
			\multirow{5}{*}{\begin{turn}{90}
					{\small\textit{From}}
				\end{turn}} 
			&  & Global & Energy & Metal & Agriculture\tabularnewline
			\cline{2-6} 
			& Global & 0.00 & 0.04 & 0.00 & 0.12 \\ 
			& Energy & 0.00 & 0.33 & 0.08 & 0.28 \\ 
			& Metal & 0.19 & 0.17 & 0.00 & 0.00 \\ 
			& Agriculture & 0.04 & 0.11 & 0.04 & 0.03 \\  
		\end{tabular}
		\ \ \ \ \ \ %
			\begin{tabular}{c|c|cccc}
				\hline 
				\multicolumn{6}{c}{{\textbf{Standard \& Poor's}}}\tabularnewline
				\hline 
				\hline 
				& \multicolumn{5}{c}{\textit{\small{}To}}\tabularnewline
				\hline 
				\multirow{5}{*}{\begin{turn}{90}
						{\small\textit{From}}
					\end{turn}} 
				&  & Global & Energy & Metal & Agriculture\tabularnewline
				\cline{2-6} 
				& Global & 0.00 & 0.17 & 0.06 & 0.12 \\ 
				& Energy & 0.00 & 0.33 & 0.00 & 0.28 \\ 
				& Metal & 0.19 & 0.25 & 0.00 & 0.00 \\ 
				& Agriculture & 0.04 & 0.11 & 0.04 & 0.03 \\  
			\end{tabular}%
	}		
	
	\bigskip
	\centering
	\resizebox{8cm}{!}{%
			\begin{tabular}{c|c|cccc}
				\hline 
				\multicolumn{6}{c}{{\textbf{Thomson Reuters}}}\tabularnewline
				\hline 
				\hline 
				& \multicolumn{5}{c}{\textit{\small{}To}}\tabularnewline
				\hline 
				\multirow{5}{*}{\begin{turn}{90}
						{\small\textit{From}}
					\end{turn}} 
				&  & Global & Energy & Metal & Agriculture\tabularnewline
				\cline{2-6} 
				& Global & 0.25 & 0.17 & 0.06 & 0.08 \\ 
				& Energy & 0.33 & 0.33 & 0.08 & 0.06 \\ 
				& Metal & 0.19 & 0.17 & 0.00 & 0.00 \\ 
				& Agriculture & 0.25 & 0.00 & 0.08 & 0.03 \\
			\end{tabular}%	
		}
\end{table}

These spillover effects might be explained by the rising importance of the ethanol and biofuel industry. In Figure \ref{Comm_Indexes_Network}, we notice that the majority of the effects among agricultural commodities involve crops that can be used for the biofuel production (like corn, wheat, soybeans), whereas it is not the case for other agriculture goods (for instance cotton). Moreover, recall that the global index for energy (ENER) presents a high out-going connectedness. Hence, we find important spillover effects from energy towards biofuel crops, in line with \cite{Chen10}; \cite{Nazlioglu12}.

Bidirectional spillovers are also observed between energy and metal in three portfolios, namely Dow Jones, Merrill Lynch and Thomson Reuters. The majority of these spillover effects involve gold  and silver, which are the most connected metal commodities (cfr. Table \ref{indexes_connect}), confirming the existence of strong dependence between energy and precious metals \citep{Sari10}.

\paragraph{Other data period}
We redo the portfolio analysis considering the period from November 2nd, 2011 to November, 1st 2013. 	
During 2013, commodity prices experienced a sharp drop mainly due to the slowdown of some emerging economies, among which China. For instance, the world corn price dropped by almost 40\% in 2013. As a result, we should not assume constant parameter values for both periods.

The commodity effect networks show major differences with the ones in Figure \ref{Comm_Indexes_Network}. The average proportion of non-zero estimated effects has almost doubled: from 6\% for the 2011-13 period to 10\% for the 2013-15 period. The lower connectedness in the period 2011-13 can be linked to the crude oil price drop that occurred in 2014, when crude oil price decreased by almost 50\%. 
As in \cite{Baffes13}, a low crude oil price could result in weaker policies sustaining the use of biofuels, hence making alternative fuel less attractive. 
As a consequence, agricultural prices drop consistently and the overall connectedness among energy and agricultural commodities increases, thus explaining the higher connectedness in the period 2013-15.

The proportion of shared non-zero effects among portfolios has also increased: from 40\% for 2011-13 to 56\% for 2013-15.
Similar to the 2013-15 period, Thomson Reuters shows the least similarity with the other portfolios (sharing on average only 10\% of its non-zero effects with the other portfolios). But in 2011-13  Credit Suise has a considerable amount of non-zero commodity effects that differ in terms of sign, direction and magnitude from the other portfolios.  
The importance of the agricultural index in the Credit Suisse network is more evident in 2011-13, it has in-going positive effects not only coming from agricultural commodities, but also from the metal and energy commodities.

Overall, we obtain different results in the two periods of analysis suggesting the presence of a structural break in commodity price dynamics. Detailed results are available from the authors upon request.

\section{Discussion \label{Discussion}}
This paper proposes the \textit{Multi-class VAR} model to analyze commodity price dynamics. 
Our method extends the literature on commodity analysis in three main directions. 
First, we study a large set of prices of different commodity types (e.g. agriculture, metal and energy). 
Second, we compare the resulting effects between markets (or investment portfolios).
We accomplish both tasks by using sparse estimation of the Multi-class VAR model, which addresses the problem of over-parameterization and allows a joint comparison between markets (or investment portfolios). 
Third, we interpret the results using networks.

% Results
We exploit the Multi-class VAR model to estimate the price effects among commodities in two data sets. 
The first one considers agricultural, metal and energy commodity prices in three different markets, namely World, India and China.
The second data set analyzes commodity prices of global, energy, metal and agriculture indexes in five different investment portfolios. 
Overall, we find few common effects between the World, Indian and Chinese markets, whereas we find evidence of more common effects among portfolios. With respect to the first data set, our results highlight the prevalence of effects involving metal goods in India and China. Moreover, we underline the existence of more spillover effects from agriculture to metal than vice versa. 
In the second data set, 
we observe outspoken spillover effects from energy towards agriculture \citep{Tyner10} and we highlight the relevance of biofuel commodities. While our modeling approach has an exploratory flavor, the majority of our results are in line with the literature on commodity price dynamics.

% Limitation
The network representation of the results brings relevant information to commodity analyst and eases the interpretation. In our application, we consider a VAR of order one, but the analysis also applies for higher order models. The choice of the VAR is in line with a large trend of the extant literature on commodity markets. 
We do acknowledge, however, that we did not include possible co-integration relations in the model. 

Further research might extend our analysis by including exogenous factors. For instance, the inclusion of interest rates might bring important insights and explain part of the overall commodity prices dynamics \citep{Akram09, Smiech15}. Another factor that might be relevant to study is the impact of excess liquidity on commodity prices. A result of the ``financialization" is the rapid increase in financial investment in commodity markets and the detachment of their trends from simple demand-offer dynamics: for instance, \cite{Belke13} provide evidence that global liquidity is one of the main drivers of commodity price. 

The Multi-class VAR might also be employed to study volatility spillover effects among a set of commodity prices. Various studies in the related literature underline the importance of the volatility analysis in commodity markets for both spot prices and derivatives \citep{Pindyck04}, with important implications in terms of risk and storage management \citep{Pindyck01}. Following \cite{Diebold15}, it is possible to incorporate the volatility spillover effects in a VAR framework. An extension of their approach to the Multi-class VAR model would take the shared volatility dynamics between markets into account.

\bigskip

\noindent
{\bf Acknowledgments.}
We gratefully acknowledge support from the FWO (Research Foundation Flanders, contract number 11N9913N) and from the GOA/12/014 project of the Research Fund KU Leuven.

\clearpage

\bibliographystyle{elsarticle-harv}
\bibliography{GFLasso_ref}

\begin{thebibliography}{47}
\expandafter\ifx\csname natexlab\endcsname\relax\def\natexlab#1{#1}\fi
\expandafter\ifx\csname url\endcsname\relax
  \def\url#1{\texttt{#1}}\fi
\expandafter\ifx\csname urlprefix\endcsname\relax\def\urlprefix{URL }\fi

\bibitem[{Achvarina and Burda(2006)}]{Achvarina06}
Achvarina, V., Burda, M., 2006. {Integration of the Chinese aluminum market
  into the global economy: Empirical case study}. Proceedings of the Third
  International Conference on Integration, University of Zagreb, Croatia, ISBN:
  953-6025-17-5.

\bibitem[{Akram(2009)}]{Akram09}
Akram, Q.~F., 2009. Commodity prices, interest rates and the dollar. Energy
  Economics 31, 838--851.

\bibitem[{Anson(2006)}]{Anson06}
Anson, M. J.~P., 2006. Handbook of Alternative Assets. Wiley, Hoboken, US.

\bibitem[{Arezki et~al.(2014)Arezki, Loungani, Van~der Ploeg, and
  Venables}]{Arezki14}
Arezki, R., Loungani, P., Van~der Ploeg, R., Venables, A., 2014. Understanding
  international commodity price fluctuations. Journal of International Money
  and Finance 42(C), 1--8.

\bibitem[{Baffes(2013)}]{Baffes13}
Baffes, J., 2013. A framework for analyzing the interplay among food, fuels,
  and biofuels. Global Food Security 2(2), 110--116.

\bibitem[{Balcombe and Rapsomanikis(2008)}]{Balcombe08}
Balcombe, K., Rapsomanikis, G., 2008. {Bayesian estimation and selection of
  nonlinear vector error correction models: The case of the sugar-ethanol-oil
  nexus in Brazi}l. American Journal of Agricultural Economics 90(3), 658--668.

\bibitem[{Belke et~al.(2013)Belke, Bordon, and Volz}]{Belke13}
Belke, A., Bordon, I.~G., Volz, U., 2013. Effects of global liquidity on
  commodity and food prices. World Development 44, 31--43.

\bibitem[{Billio et~al.(2012)Billio, Getmansky, Lo, and Pelizzon}]{Billio12}
Billio, M., Getmansky, M., Lo, A.~W., Pelizzon, L., 2012. Econometric measures
  of connectedness and systemic risk in the finance and insurance sectors.
  Journal of Financial Economics 104(3), 535--559.

\bibitem[{Bukenya and Labys(2005)}]{Bukenya05}
Bukenya, J.~O., Labys, W.~C., 2005. Price convergence on world commodity
  markets: Fact or fiction? International Regional Science Review 28(3),
  302--329.

\bibitem[{Cashin and McDermott(2002)}]{Cashin02}
Cashin, P., McDermott, C., 2002. The long-run behavior of commodity prices:
  Small trends and big variability. IMF Staff Papers 49(2), 175--199.

\bibitem[{Chen(2015)}]{Chen15}
Chen, P., 2015. {Global oil prices, macroeconomic fundamentals and China's
  commodity sector comovements}. Energy Policy 87, 284--294.

\bibitem[{Chen et~al.(2010)Chen, Kuo, and Chen}]{Chen10}
Chen, S.-T., Kuo, H.-I., Chen, C.-C., 2010. Modeling the relationship between
  the oil price and global food prices. Applied Energy 87(8), 2517--2525.

\bibitem[{Chen et~al.(2012)Chen, Lin, Kim, Carbonell, and Xing}]{Chen12}
Chen, X., Lin, Q., Kim, S., Carbonell, J.~G., Xing, E.~P., 2012. Smoothing
  proximal gradient method for general structured sparse regression. The Annals
  of Applied Statistics 6(2), 719--752.

\bibitem[{Danaher et~al.(2014)Danaher, Wang, and Witten}]{Danaher14}
Danaher, P., Wang, P., Witten, D.~M., 2014. The joint graphical lasso for
  inverse covariance estimation across multiple classes. Journal of the Royal
  Statistical Society Series B 76(2), 373--397.

\bibitem[{Deaton(1999)}]{Deaton99}
Deaton, A., 1999. {Commodity prices and growth in Africa}. Journal of Economic
  Perspectives 13(3), 23--40.

\bibitem[{Diebold and Yilmaz(2015)}]{Diebold15}
Diebold, F.~X., Yilmaz, K., 2015. Financial and macroeconomics connectedness: A
  network approach to measurement and monitoring. Oxford University Press, New
  York, US.

\bibitem[{Franckel and Rose(2010)}]{Franckel10}
Franckel, J.~A., Rose, A.~K., 2010. Determinants of agricultural and mineral
  commodity prices. In: Fry, R. Jones, C. Kent, C. Inflation in an era of
  relative price shocks. Reserve Bank of Australia.

\bibitem[{Hassouneh et~al.(2012)Hassouneh, Serra, Goodwin, and
  Gil}]{Hassouneh12}
Hassouneh, I., Serra, T., Goodwin, B.~K., Gil, J.~M., 2012. Non-parametric and
  parametric modeling of biodiesel, sunflower oil, and crude oil price
  relationships. Energy Economics 34(5), 15078--1513.

\bibitem[{Isard(1977)}]{Isard77}
Isard, P., 1977. How far can we push the law of one price? The American
  Economic Review 67(5), 942--948.

\bibitem[{Jain and Ghosh(2013)}]{Jain13}
Jain, A., Ghosh, S., 2013. {Dynamics of global oil prices, exchange rate and
  precious metal prices in India}. Resources Policy 38(1), 88--93.

\bibitem[{Jenatton et~al.(2011)Jenatton, Audibert, and Bach}]{Jenatton11}
Jenatton, R., Audibert, J.~Y., Bach, F., 2011. Structured variable selection
  with sparsity-inducing norms. Journal of Machine Learning Research 12,
  2777--2824.

\bibitem[{Kim and Xing(2009)}]{Kim09}
Kim, S., Xing, E.~P., 2009. Statistical estimation of correlated genome
  associations to a quantitative trait network. PLoS Genetics 5(8): e1000587.
  doi:10.1371/ journal.pgen.1000587.

\bibitem[{Klotz et~al.(2014)Klotz, Lin, and Hsu}]{Klotz14}
Klotz, P., Lin, T.~C., Hsu, S.~H., 2014. {Global commodity prices, economic
  activity and monetary policy: The relevance of China}. Resources Policy 42,
  1--9.

\bibitem[{Kolaczyk(2009)}]{Kolaczyk09}
Kolaczyk, E.~D., 2009. Statistical analysis of network data: Methods and
  models. Springer, New York.

\bibitem[{Labys(2006)}]{Labys06}
Labys, W., 2006. Modeling and forecasting primary commodity prices. Ashgate
  Publishing, Chippenham, UK.

\bibitem[{Levin et~al.(2002)Levin, Lin, and Chu}]{Levin02}
Levin, A., Lin, C.-F., Chu, C.-S.~J., 2002. Unit root tests in panel data:
  Asymptotic and finite-sample properties. Journal of Econometrics 108(1),
  1--24.

\bibitem[{Nazlioglu et~al.(2013)Nazlioglu, Erdem, and Soytas}]{Nazlioglu13}
Nazlioglu, S., Erdem, C., Soytas, U., 2013. Volatility spillover between oil
  and agricultural commodity markets. Energy Economics 36, 658--665.

\bibitem[{Nazlioglu and Soyatas(2012)}]{Nazlioglu12}
Nazlioglu, S., Soyatas, U., 2012. {Oil price, agricultural commodity prices,
  and the dollar: A panel cointegration and causality analysis}. Energy
  Economics 34(4), 1098--1104.

\bibitem[{Pindyck(2001)}]{Pindyck01}
Pindyck, R.~S., 2001. The dynamics of commodity spot and futures markets: A
  primer. The Energy Journal 22(3), 1--29.

\bibitem[{Pindyck(2004)}]{Pindyck04}
Pindyck, R.~S., 2004. Volatility and commodity price dynamics. The Journal of
  Futures Markets 24(11), 1029--1047.

\bibitem[{Pindyck and Rotemberg(1990)}]{Pindyck90}
Pindyck, R.~S., Rotemberg, J.~J., 1990. The excess co-movement of commodity
  prices. The Economic Journal 100(403), 1173--1189.

\bibitem[{Pitfield et~al.(2010)Pitfield, Brown, and Idoine}]{Pitfield10}
Pitfield, P. E.~J., Brown, T.~J., Idoine, N.~E., 2010. Mineral information and
  statistics for the {BRIC} countries 1999-2008. British Geological Survey,
  Keyworth, Nottingham.

\bibitem[{Rapsomanikis(2011)}]{Rapsomanikis11}
Rapsomanikis, G., 2011. Price transmission and volatility spillovers in food
  markets. In: Prakash, A. Safeguarding food security in volatile global
  markets. Food and Agriculture Organization of the United Nations, Rome.

\bibitem[{Ravaillon(1986)}]{Ravallion86}
Ravaillon, M., 1986. Testing market integration. American Journal of
  Agricultural Economics 68(1), 102--109.

\bibitem[{Rezitis(2015)}]{Rezitis15}
Rezitis, A.~N., 2015. The relationship between agricultural commodity prices,
  crude oil prices and {US} dollar exchange rates: A panel {VAR} approach and
  causality analysis. International Review of Applied Economics 29(3),
  403--434.

\bibitem[{Rossen(2015)}]{Rossen15}
Rossen, A., 2015. {What are metal prices like? Co-movement, price cycles and
  long-run trends}. Resources Policy 45, 255--276.

\bibitem[{Rothman et~al.(2010)Rothman, Levina, and Zhu}]{Rothman10}
Rothman, A.~J., Levina, E., Zhu, J., 2010. Sparse multivariate regression with
  covariance estimation. Journal of Computational and Graphical Statistics 19
  (4), 947--962.

\bibitem[{Sari et~al.(2010)Sari, Hammaoudeh, and Soytas}]{Sari10}
Sari, R., Hammaoudeh, S., Soytas, U., 2010. Dynamics of oil price, precious
  metal prices, and exchange rate. Energy Economics 32(2), 351--362.

\bibitem[{Serra et~al.(2011)Serra, Ziberman, Gil, and Goodwin}]{Serra11mean}
Serra, T., Ziberman, D., Gil, J.~M., Goodwin, B.~K., 2011. {Nonlinearities in
  the U.S. corn-ethanol-oil-gasoline price system}. Agricultural Economics
  42(1), 34--45.

\bibitem[{Serra and Zilberman(2013)}]{Serra13}
Serra, T., Zilberman, D., 2013. {Biofuel-related price transmission literature:
  A review}. Energy Economics 37, 141--151.

\bibitem[{Smiech et~al.(2015)Smiech, Papiez, and Dabrowski}]{Smiech15}
Smiech, S., Papiez, M., Dabrowski, M.~A., 2015. {Does the euro area
  macroeconomicely affect global commidity prices? Evidence from a SVAR
  approach}. International Review of Economics 39, 485--503.

\bibitem[{Tibshirani(1996)}]{Tibshirani96}
Tibshirani, R., 1996. Regression shrinkage and selection via the lasso. Journal
  of the Royal Statistical Society Series B 58(1), 267--288.

\bibitem[{Tibshirani et~al.(2012)Tibshirani, Bien, Friedman, and
  Hastie}]{Tibshirani12}
Tibshirani, R., Bien, J., Friedman, J., Hastie, T., 2012. {Strong rules for
  discarding predictors in lasso-type problems}. Journal of the Royal
  Statistical Society Series B 74(2), 245--266.

\bibitem[{Tibshirani et~al.(2005)Tibshirani, Saunders, Rosset, Zhu, and
  Knight}]{Tibshirani05}
Tibshirani, R., Saunders, M., Rosset, S., Zhu, J., Knight, K., 2005. Sparsity
  and smoothness via the fused lasso. Journal of the Royal Statistical Society
  Series B 67(1), 91--108.

\bibitem[{Tyner(2010)}]{Tyner10}
Tyner, W.~E., 2010. The integration of energy and agricultural markets.
  Agricultural Economics 41(1), 193--201.

\bibitem[{Wainwright(2014)}]{Wainwright14}
Wainwright, M.~J., 2014. Structured regularizers for high-dimensional problems:
  Statistical and computational issues. Annual Review of Statistics and its
  Application 1, 233--253.

\bibitem[{Yang et~al.(2000)Yang, Bessler, and Leatham}]{Yang00}
Yang, J., Bessler, D.~A., Leatham, D.~J., 2000. {The law of one price:
  Developed and developing country market integration}. Journal of Agricultural
  and Applied Economics 32(3), 429--440.

\end{thebibliography}

\end{document}